\documentclass[aps,preprint,floats,epsf,epsfig,nofootinbib,letter]{revtex4}
\usepackage{epsfig}
\usepackage{graphicx}
\usepackage{dcolumn}
\usepackage{bm}
\usepackage{subfigure}
\usepackage{hyperref}                 
\usepackage{amsmath}
\usepackage{amsfonts}
\usepackage{xcolor}
\usepackage{soul}

\begin{document}
\def\be{\begin{eqnarray}}
\def\en{\end{eqnarray}}
\def\non{\nonumber}
\def\la{\langle}
\def\ra{\rangle}
\def\ov{\overline}
\def\Br{{\mathcal B}}
\def\A{{\mathcal A}}
\def\B{{\cal B}}
\def\D{{\cal D}}
\def\Bbar{\overline{\cal B}}
\def\bfB{{\rm\bf B}}
\def\bfBB{{\rm\bf B}\overline{\rm\bf B}}
\def\bfBcBc{{\rm\bf B}_c\overline{\rm\bf B}_c}
\def\BB{{{\cal B}_c \overline {\cal B}_c}}
\def\BD{{{\cal B} \overline {\cal D}}}
\def\DB{{{\cal D} \overline {\cal B}}}
\def\DD{{{\cal D} \overline {\cal D}}}
\def\sq{\sqrt}


\title{Revisiting two-body charmed anti-charmed baryonic $B$ decays}

\author{Chun-Khiang Chua}
\affiliation{Department of Physics and Center for High Energy Physics,
Chung Yuan Christian University,
Chung-Li, Taiwan 320, Republic of China}

\date{\today}

\begin{abstract}
The study of two-body charmed anti-charmed baryonic $\overline B\to {\cal B}_c \overline {\cal B}_c$ decays using the topological amplitude approach is revisited.
We include the $b\to u\bar u q$ contributions, in addition to the $b\to c\bar c q$ contributions, to these decays.
The topological decomposition of $\overline B\to {\cal B}_c(\bf {\bar 3_f}) \overline {\cal B}_c(\bf { 3_f})$,
${\cal B}_c(\bf 6_f) \overline {\cal B}_c(\bf { 3_f})$,
${\cal B}_c(\bf {\bar 3_f}) \overline {\cal B}_c(\bf {\bar 6_f})$
and
${\cal B}_c(\bf 6_f) \overline {\cal B}_c(\bf {\bar 6_f})$ decay amplitudes are updated accordingly.
Although the $b\to u\bar u q$ contributions are Cabibbo-Kobayashi-Maskawa (CKM) suppressed in $\Delta S=-1$ transitions, their effects are significantly amplified in $\Delta S=0$ transitions, 
as the relative size of the CKM factors in $b\to u\bar u q$ and $b\to c\bar c q$ contributions in $\Delta S=0$ transitions is enlarged by $\lambda^{-2}$, roughly a factor of 20, from the one in $\Delta S=-1$ transitions.
Present data in $\overline B\to {\cal B}_c(\bf {\bar 3_f}) \overline {\cal B}_c(\bf { 3_f})$ decay rates allow or even slightly prefer non-negligible $b\to u\bar u q$ contributions. 
These $b\to u\bar u q$ contributions can allow much larger rates for $B^-\to \Xi_c^0 \bar \Xi_c^{-}$,
$\overline B{}_s^0\to \Lambda_c^+ \bar \Xi_c^{-}$
and                                      
$\overline B{}^0\to \Xi_c^+ \bar \Xi_c^{-}$ decays than those only have $b\to c\bar c d$ contributions. 
Several modes containing only $b\to u\bar u q$ contributions in $\overline B\to {\cal B}_c(\bf 6_f) \overline {\cal B}_c(\bf { 3_f})$,
${\cal B}_c(\bf {\bar 3_f}) \overline {\cal B}_c(\bf {\bar 6_f})$ decays are also identified.
Measuring these decays will be interesting and useful in clarifying the role of $b\to u\bar u q$ contributions in two-body charmed anti-charmed baryonic $\overline B$ decays.

\end{abstract}


\maketitle


\vfill\eject

\section{Introduction}

There is continuous experimental progress in investigating two-body charmed anti-charmed baryonic $B$ decays. 
For example, recently, LHCb reported  $\bar B^0\to \Lambda^+_c\bar \Lambda^-_c$ and $\bar B^0_s\to \Lambda^+_c\bar \Lambda^-_c$ decays \cite{LHCb:2025ueu},
while Belle II reported
$B^-\to \Xi_c^+\bar\Sigma_c(2455)^{--}$, $\bar B^0\to \Xi_c^0\bar\Sigma_c(2455)^0$ decays \cite{Belle:2025nup},
and $B^-\to \Xi_c^{\prime +}\bar\Sigma_c(2455)^{--}$,  $\bar B^0\to \Xi_c^{\prime 0}\bar\Sigma_c(2455)^0 $ decays \cite{Belle-II:2026ynf}.
See Table~\ref{tab: expt} for a summary of the current situation~\cite{ParticleDataGroup:2024cfk,
Belle:2018kzz,
Belle:2019pze,
Belle:2025nup,
LHCb:2025ueu,
Belle:2019bgi,
Belle-II:2026ynf}.
It is timely to study these decays.

\begin{table}[b!]
\caption{\label{tab: expt}
Current experimental results of $\bar B_{u,d,s}\to\B_c\overline \B_c$ branching ratios (in the unit of $10^{-4}$). 
}
\begin{ruledtabular}
\begin{tabular}{lcccr}
Mode
          & LHCb
          & Belle/ Belle-II
          & PDG~\cite{ParticleDataGroup:2024cfk}
          \\
\hline $B^-\to\Xi_c^0\bar \Lambda_c^-$
          & 
          & $9.51\pm2.10\pm0.88$ \cite{Belle:2018kzz}
          & $9.5\pm2.3$
          \\
$B^-\to\Xi_c^{\prime 0}\bar \Lambda_c^-$
          & 
          & $3.4\pm 2.0$ \cite{Belle:2019pze}
          & $<6.5$
          \\
$B^-\to\Xi_c^{0}(2645)\bar \Lambda_c^-$
          & 
          & $4.4\pm 2.4$ \cite{Belle:2019pze}
          & $<7.9$
          \\
$B^-\to\Xi_c^{0}(2790)\bar \Lambda_c^-$
          & 
          & $1.1\pm 0.4$ \cite{Belle:2019pze}
          & $1.1\pm 0.4$
          \\
$B^-\to \Xi_c^+\bar\Sigma_c(2455)^{--} $
          & 
          & $5.74\pm1.11\pm0.42^{+2.47}_{-1.53}$ \cite{Belle:2025nup}
          & $5.7^{+2.7}_{-1.9}$ 
          \\          
$B^-\to \Xi_c^{\prime +}\bar\Sigma_c(2455)^{--} $
          & 
          & $16.8\pm3.1\pm 1.2^{+14.9}_{-5.4}$ \cite{Belle-II:2026ynf}
          & 
          \\     
\hline $\bar B^0\to\Xi_c^+\bar \Lambda_c^-$
          & 
          & $11.6\pm4.2\pm1.5$ \cite{Belle:2019bgi}
          & $11\pm 8$
          \\
$\bar B^0\to \Lambda^+_c\bar \Lambda^-_c$
          & $0.101^{+0.027}_{-0.028}\pm0.008\pm0.015$ \cite{LHCb:2025ueu}
          &  
          & $<0.16$
          \\
$\bar B^0\to \Xi_c^0\bar\Sigma_c(2455)^0 $
          & 
          & $4.83\pm1.12\pm0.37^{+0.72}_{-0.60}$ \cite{Belle:2025nup}
          & $4.8\pm 1.4$
          \\        
$\bar B^0\to \Xi_c^{\prime 0}\bar\Sigma_c(2455)^0 $
          & 
          & $12.8\pm 3.2\pm 1.0^{+3.0}_{-2.1}$ \cite{Belle-II:2026ynf}
          & 
          \\  
\hline $\bar B^0_s\to \Lambda^+_c\bar \Lambda^-_c$
          & $0.50\pm0.13\pm0.05\pm0.08$   \cite{LHCb:2025ueu}
          &
          & $<0.8$
          \\                                                         
\end{tabular}
\end{ruledtabular}
\end{table}

\begin{table}[t!]
\caption{\label{tab: Br BtoBcBcbar 3bar3 0}
Decay amplitudes and branching ratios (in the unit of $10^{-4}$) of $B^-\to \Xi_c^0 \bar \Lambda_c^{-}$ and $\overline B{}^0\to \Xi_c^+ \bar \Lambda_c^{-}$ decays.
Theoretical results are taken from refs. \cite{Hsiao:2023mud} and \cite{Chua:2026awd}.  
}
\footnotesize{
\begin{ruledtabular}
\begin{tabular}{lcccc}
Mode
          & $A \left(\overline B_q\to \B_c({\bf \bar 3_f})\overline \B_c ({\bf 3_f})\right)$
          & $Br \left(\overline B_q\to \B_c({\bf \bar 3_f})\overline \B_c ({\bf 3_f})\right)$
          & $Br \left(\overline B_q\to \B_c({\bf \bar 3_f})\overline \B_c ({\bf 3_f})\right)$
          & Expt.
           \\
          & ref. \cite{Chua:2026awd}
          & ref. \cite{Hsiao:2023mud}
          & ref. \cite{Chua:2026awd}
          & 
           \\
\hline
$B^-\to \Xi_c^0 \bar \Lambda_c^{-}$
          & $C_{1,v}$  
          & $7.8^{+2.3}_{-2.0}$
          & $10.06_{-2.04}^{+1.98}$
          & $9.51\pm 2.28$~\cite{Belle:2018kzz}
          \\
$\overline B{}^0\to \Xi_c^+ \bar \Lambda_c^{-}$
          & $C_{1,v}$
          & $7.2^{+2.1}_{-1.9}$
          & $9.34_{-1.90}^{+1.84}$
         & $11.6\pm 4.46$~\cite{Belle:2019bgi}
          \\
\end{tabular}
\end{ruledtabular}
}
\end{table}

\begin{figure}[t]
\centering
 \subfigure[]{
  \includegraphics[width=0.47\textwidth]{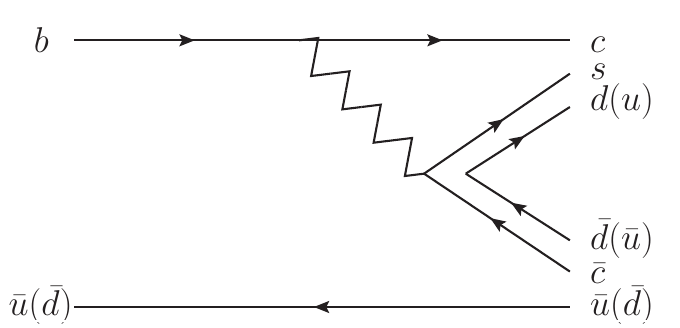}
}
\hspace{12pt}
\subfigure[]{
  \includegraphics[width=0.45\textwidth]{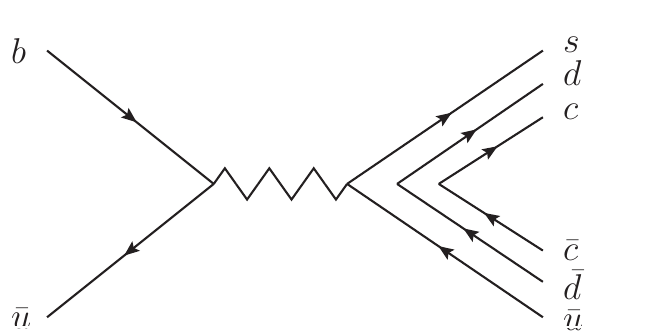}
}
\\
 \subfigure[]{
  \includegraphics[width=0.47\textwidth]{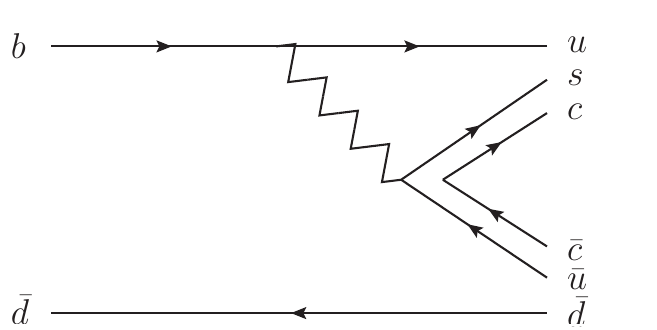}
}
\caption{Topological diagrams of (a) $b\to c\bar c s$ transition and (b), (c) $b\to u\bar us$ transition contributing to $B^-\to \Xi_c^0 \bar \Lambda_c^{-}$ and $\overline B{}^0\to \Xi_c^+ \bar \Lambda_c^{-}$ decays.
} 
\label{fig:TA XiLambda}
\end{figure}

There are many theoretical studies on $\overline B\to\B_c\overline \B_c$ decays, see
\cite{Chernyak:1990ag, Ball:1990fw,Cheng:2005vd,Chen:2006fsa,Cheng:2009yz,Hsiao:2023mud,Rui:2024xgc,Geng:2025yna, Chua:2026awd,Duan:2026gly}.
For example, 
the $\overline B\to {\cal B}_c(\bf {\bar 3_f}) \overline {\cal B}_c(\bf { 3_f})$ decays, where ${\cal B}_c(\bf {\bar 3_f})$ stands for an anti-triplet charmed baryon,
are studied using topological amplitudes in refs. \cite{Hsiao:2023mud, Chua:2026awd}.
It is known that $B^-\to \Xi_c^0 \bar \Lambda_c^{-}$ and $\overline B{}^0\to \Xi_c^+ \bar \Lambda_c^{-}$ decays are SU(2) related modes.
As shown in Table~\ref{tab: Br BtoBcBcbar 3bar3 0}, they have identical amplitudes, namely, the internal $W$-tree amplitude, 
and should have identical decay rates in the SU(2) limit.
Both ref.\cite{Hsiao:2023mud} and ref.\cite{Chua:2026awd} obtain a slightly larger $B^-\to \Xi_c^0 \bar \Lambda_c^{-}$ branching ratio comparing to the branching ratio of $\overline B{}^0\to \Xi_c^+ \bar \Lambda_c^{-}$ decay,
reflecting the differences in $B^-$ and $\bar B^0$ lifetimes, see Table~\ref{tab: Br BtoBcBcbar 3bar3 0}.
However, the central values of the data seem to point in the opposite direction, leaving room for contributions from some SU(2) breaking effects.

Indeed, in $B^-\to \Xi_c^0 \bar \Lambda_c^{-}$ and $\overline B{}^0\to \Xi_c^+ \bar \Lambda_c^{-}$ decays, only $b\to c\,\bar c s$ transition, see Fig.~\ref{fig:TA XiLambda}~(a), were considered.
However, as shown in Fig.~\ref{fig:TA XiLambda} (b) and (c), diagrams from the $b\to u \bar u s$ transition, which is an SU(2) breaking transition, 
can also contribute to these decays.
With these contributions, the amplitudes of $B^-\to \Xi_c^0 \bar \Lambda_c^{-}$ and $\overline B{}^0\to \Xi_c^+ \bar \Lambda_c^{-}$ decays are no longer identical. 
Note that, so far, systematic studies of these $b\to u \,\bar u q$ contributions have not been considered.

 It is true that these $b\to u \bar u s$ transition diagrams are sub-leading in the $\Delta S=-1$ transition modes,
 but their contributions are likely to surface in $\Delta S=0$ transition modes.  
 The argument is as follows.
The decay amplitudes in $\Delta S=-1$ and $\Delta S=0$ transitions have the following structure,
\be
A(\Delta S=-1)&=& A^c+A^u=V_{cb} V^*_{cs} \,a^c+V_{ub} V^*_{us} \,a^u,
\non\\
A'(\Delta S=0)&=& A^{\prime c}+A^{\prime u}=V_{cb} V^*_{cd} \,a^c+V_{ub} V^*_{ud} \,a^u,
\en
where $V_{cb}$, $V_{cs(d)}$, $V_{ub}$ and $V_{us(d)}$ are Cabibbo-Kobayashi-Maskawa (CKM) matrix elements.
The $A^{(\prime) c}=V_{cb} V^*_{cs(d)} \,a^c$ part is from $b\to c\bar c q$ contributions, while the $A^{(\prime) u}=V_{ub} V^*_{us(d)} \,a^u$ part is from $b\to u\bar u q$ contributions.
Note that naively the ratio of the sizes of $b\to u\bar u q$ and $b\to c\bar c q$ terms in $\Delta S=0$ and $-1$ transitions 
is given by
\be
\left|\frac{A^{\prime u}}{A^{\prime c}}\right|
\bigg/
\left|\frac{A^u}{A^c}\right|
=
\left|\frac{V_{ub} V^*_{ud} \, a^u}{V_{cb} V^*_{cd} \, a^c}\right|
\bigg/
\left|\frac{V_{ub} V^*_{us} \, a^u}{V_{cb} V^*_{cs} \,a^c}\right|
=
\left|\frac{V_{ub} V^*_{ud}}{V_{cb} V^*_{cd}}\right|
\bigg/
\left|\frac{V_{ub} V^*_{us}}{V_{cb} V^*_{cs}}\right|
\simeq \frac{1}{\lambda^2}
\simeq 20.
\label{eq: 20}
\en
For simplicity, we cancel out the $a^u/a^c$ ratio in the above equation. 
In any case, the CKM double ratio is almost always there. 
Consequently, the relative sizes of $b\to u\, \bar u q$ and $b\to c\, \bar c q$ contributions are enhanced by roughly 20 times in the $\Delta S=0$ transition compared to those in the $\Delta S=-1$ transition.
Hence, the effect of the $b\to u\, \bar u q$ transition, relative to the $b\to c\, \bar c q$ transition, is amplified significantly in the $\Delta S=0$ transition.
It will be interesting to investigate their effects in two-body charmed anti-charmed baryonic $\overline B\to {\cal B}_c \overline {\cal B}_c$ decays.

The structure of this paper is as follows. 
In Sec.~II, updated topological amplitudes of all $\overline B\to {\cal B}_c \overline {\cal B}_c$ decays by including $b\to u\bar u q$ contributions are given. 
Numerical results on rates of $\overline B\to {\cal B}_c(\bf {\bar 3_f}) \overline {\cal B}_c(\bf { 3_f})$ are shown in Sec.~III,
and we also comment on other modes.
Conclusions will be given in Sec. V.

\section{Updating topological amplitudes of $\overline B\to {\cal B}_c \overline {\cal B}_c$ decays}

\subsection{Formalism}

We follow ref. \cite{Chua:2026awd} to decompose the decay amplitudes in terms of topological amplitudes, 
but including contributions from $b\to u\bar u q$ transitions in addition to $b\to c\bar c q$ contributions.

\begin{figure}[t]
\centering
 \subfigure[]{
  \includegraphics[width=0.45\textwidth]{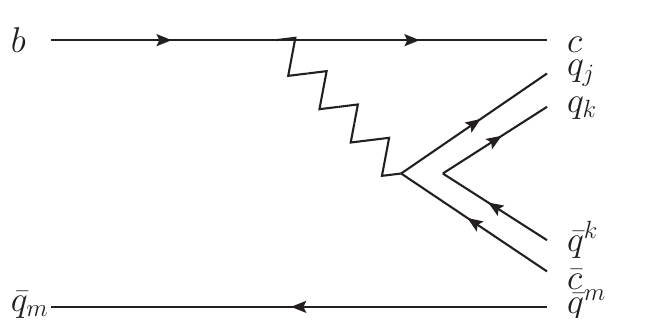}
}
\hspace{12pt}
\subfigure[]{
  \includegraphics[width=0.45\textwidth]{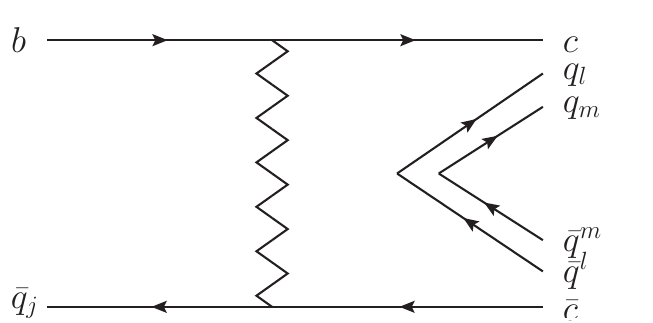}
}
\\
 \subfigure[]{
  \includegraphics[width=0.45\textwidth]{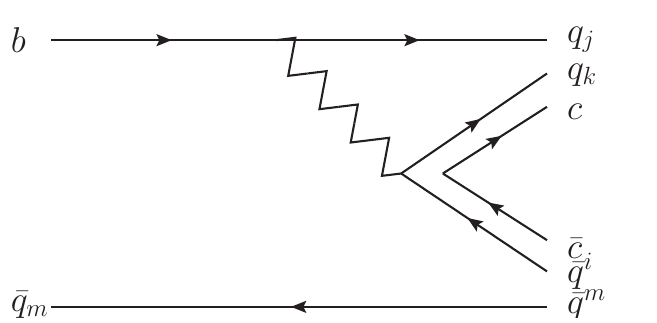}
}
\hspace{12pt}
\subfigure[]{
  \includegraphics[width=0.45\textwidth]{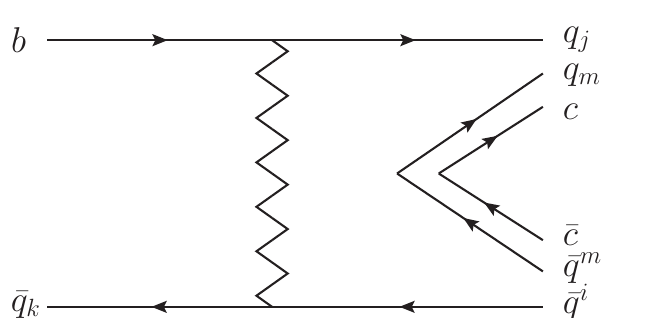}
}
\\
 \subfigure[]{
  \includegraphics[width=0.45\textwidth]{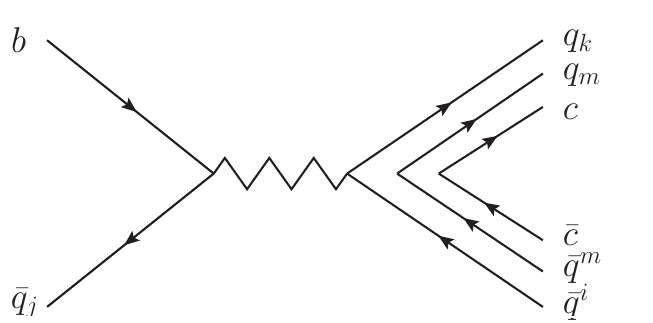}
}
\caption{Topological diagrams of 
  (a) $C$ (internal $W$-tree) and (b) $E$ ($W$-exchange) amplitudes from $b\to c \,\bar c q$ transition
  and 
   (c) $D$ (internal $W$-tree), (d) $F$ ($W$-exchange) and (e) $A$ (annihilation) amplitudes from $b\to u \,\bar u q$ transition
   in $\overline B$ to charmed baryon pair decays. 
  These are flavor flow diagrams. 
  Diagrams (c)-(e) are new.
} \label{fig:TA}
\end{figure}

In $\Delta S=-1$ $\overline B \to \B_c \overline \B_c $ decays,
the $b\to c \bar c s$ transition
is governed by a $(\bar c b)_{V-A} (\bar s c)_{V-A}$ operator,
while the $b\to u \bar u s$ transition
is governed by a $(\bar u b)_{V-A} (\bar s u)_{V-A}$ operator.
These operators have the following structures,
\begin{eqnarray}
&&(\bar c b )(\bar s c )
 =H^i (\bar c b) (\bar q_i c),
\non\\
&&
(\bar u b )(\bar s u )
 =H^{jk}_i (\bar q_j b) (\bar q_k q^i),
\non\\
&&H^{13}_1=1=H^3, 
\quad
{\rm otherwise}\,\,\,H^{ik}_j=H^k=0,
\label{eq: Hjki Hk}
\end{eqnarray}
respectively.
Note that the above equations also apply to the $\Delta S=0$
case, but with $s$ and $H^{13}_1=1=H^3$ 
replaced by 
$d$ and $H'^{12}_1=1=H'^2$, respectively.
Consequently, the Hamiltonians governing various $\Delta S=-1$ $\overline B \to \B_c \overline \B_c $ decays are given by
\be
H_{\rm eff}\left(\overline B \to \B_c(\bf {\bar 3_f}) \overline \B_c(\bf { 3_f}) \right)
&=& 
       C_1 \, \overline B_m H^j \overline \B_{c [jk]}  \B_c^{[km]}
       -D_1\,\overline B_m H^{jk}_i \overline \B_{c [jk]}  \B_c^{[im]}
\non\\       
&&      +\frac{1}{2}E_1 \, \overline B_j H^j \overline \B_{c [lm]}  \B_c^{[ml]}
     +F_1 \overline B_k H^{jk}_i \overline \B_{c [jm]}  \B_c^{[mi]}
\non\\
&&
       +A_1 \overline B_j H^{jk}_i \overline \B_{c [km]}  \B_c^{[mi]},
 \non\\
H_{\rm eff}\left(\overline B \to \B_c(\bf 6_f) \overline \B_c(\bf { 3_f}) \right)
&=& 
       C_2 \, \overline B_m H^j \overline \B_{c\{jk\}}  \B_c^{[km]}
     +D_2\,\overline B_m H^{jk}_i \overline \B_{c \{jk\}}  \B_c^{[im]}
 \non\\
&& -F_2 \overline B_k H^{jk}_i \overline \B_{c \{jm\}}  \B_c^{[mi]}
   +A_2 \overline B_j H^{jk}_i \overline \B_{c \{km\}}  \B_c^{[mi]},
\non\\
H_{\rm eff}\left(\overline B \to \B_c(\bf {\bar 3_f}) \overline \B_c(\bf {\bar 6_f}) \right)
&=& 
       C_3 \, \overline B_m H^j \overline \B_{c[jk]}  \B_c^{\{km\}}
       -D_3\,\overline B_m H^{jk}_i \overline \B_{c [jk]}  \B_c^{\{im\}}
 \non\\
&&     -F_3 \overline B_k H^{jk}_i \overline \B_{c [jm]}  \B_c^{\{mi\}}
          +A_3 \overline B_j H^{jk}_i \overline \B_{c [km]}  \B_c^{\{mi\}},
 \non\\
H_{\rm eff}\left(\overline B \to \B_c(\bf 6_f) \overline \B_c(\bf {\bar 6_f}) \right)
&=& 
       C_4 \, \overline B_m H^j \overline \B_{c\{jk\}}  \B_c^{\{km\}}
       +\frac{1}{2}E_4 \, \overline B_j H^j \overline \B_{c\{lm\}}  \B_c^{\{ml\}}
\non\\
 &&+D_4\,\overline B_m H^{jk}_i \overline \B_{c \{jk\}}  \B_c^{\{im\}}
     +F_4 \overline B_k H^{jk}_i \overline \B_{c \{jm\}}  \B_c^{\{mi\}}
 \non\\
&& 
      +A_4 \overline B_j H^{jk}_i \overline \B_{c \{km\}}  \B_c^{\{mi\}}.
\label{eq: H0}
\en
Some explanations are required.
In the above equation,
we have
\be
&&
\quad
\B_c^{\{11\}}=\sqrt2 \Sigma^{++}_c,\,\,
\B_c^{\{12\}}=\B_c^{\{21\}}=\Sigma^{+}_c,\,\,
\B_c^{\{22\}}=\sqrt2 \Sigma^0_c,
\non\\
&&
\B_c^{\{13\}}=\B_c^{\{31\}}=\Xi^{\prime +}_c,\,\,
\B_c^{\{23\}}=\B_c^{\{32\}}=\Xi^{\prime 0}_c,\,\,
\B_c^{\{33\}}=\sqrt2 \Omega^0_c,
\en
for sextet ($\bf 6_f$) charmed baryons,
while for anti-triplet $({\bf \bar 3_f})$ charmed baryons, we have
\be
\B_c^{[12]}=-\B_c^{[21]}=\Lambda^+_c,\,\,
\B_c^{[23]}=-\B_c^{[32]}=\Xi^0_c,\,\,
\B_c^{[31]}=-\B_c^{[13]}=\Xi^+_c.
\en
Some signs in Eq. (\ref{eq: H0}) are introduced for later purposes.
$C$ and $E$ amplitudes denote internal $W$-tree and $W$-exchange amplitudes from $b\to c\bar c s$ transition, respectively, see Fig. \ref{fig:TA} (a) and (b),
while $D$, $F$ and $A$ are internal $W$-tree, $W$-exchange and annihilation diagrams from $b\to u\bar u s$ transition, respectively, see Fig. \ref{fig:TA} (c)--(e).
The latter are new in this work.
Following refs. \cite{Hsiao:2023mud, Rui:2024xgc}, we do not include penguin amplitudes, as their contributions to averaged rates are expected to be subleading and they do not introduce isospin-breaking contributions. 

Similarly, the Hamiltonians governing various $\Delta S=0$ $\overline B \to \B_c \overline \B_c $ decays are given by
\be
H_{\rm eff}\left(\overline B \to \B_c(\bf {\bar 3_f}) \overline \B_c(\bf { 3_f}) \right)
&=& 
       C'_1 \, \overline B_m H'^j \overline \B_{c [jk]}  \B_c^{[km]}
      +\frac{1}{2} E'_1 \, \overline B_j H'^j \overline \B_{c [lm]}  \B_c^{[ml]}
 \non\\
 &&-D'_1\,\overline B_m H'^{jk}_i \overline \B_{c [jk]}  \B_c^{[im]}
     +F'_1 \overline B_k H'^{jk}_i \overline \B_{c [jm]}  \B_c^{[mi]}
 \non\\
&& +A'_1 \overline B_j H'^{jk}_i \overline \B_{c [km]}  \B_c^{[mi]},
 \en
 \be
H_{\rm eff}\left(\overline B \to \B_c(\bf 6_f) \overline \B_c(\bf { 3_f}) \right)
&=& 
       C'_2 \, \overline B_m H'^j \overline \B_{c\{jk\}}  \B_c^{[km]}
     +D'_2\,\overline B_m H'^{jk}_i \overline \B_{c \{jk\}}  \B_c^{[im]}
 \non\\
&& -F'_2 \overline B_k H'^{jk}_i \overline \B_{c \{jm\}}  \B_c^{[mi]}
    +A'_2 \overline B_j H'^{jk}_i \overline \B_{c \{km\}}  \B_c^{[mi]},
\en
\be
H_{\rm eff}\left(\overline B \to \B_c(\bf {\bar 3_f}) \overline \B_c(\bf {\bar 6_f}) \right)
&=& 
       C'_3 \, \overline B_m 'H^j \overline \B_{c[jk]}  \B_c^{\{km\}}
       -D'_3\,\overline B_m H'^{jk}_i \overline \B_{c [jk]}  \B_c^{\{im\}}
 \non\\
&&     -F'_3 \overline B_k H'^{jk}_i \overline \B_{c [jm]}  \B_c^{\{mi\}}
 +A'_3 \overline B_j H'^{jk}_i \overline \B_{c [km]}  \B_c^{\{mi\}},
 \en
 \be
H_{\rm eff}\left(\overline B \to \B_c(\bf 6_f) \overline \B_c(\bf {\bar 6_f}) \right)
&=& 
       C'_4 \, \overline B_m H'^j \overline \B_{c\{jk\}}  \B_c^{\{km\}}
      +\frac{1}{2}E'_4 \, \overline B_j H^j \overline \B_{c\{lm\}}  \B_c^{\{ml\}}
\non\\
 &&+D'_4\,\overline B_m H'^{jk}_i \overline \B_{c \{jk\}}  \B_c^{\{im\}}
     +F'_4 \overline B_k H'^{jk}_i \overline \B_{c \{jm\}}  \B_c^{\{mi\}}
 \non\\
&& +A'_4 \overline B_j H'^{jk}_i \overline \B_{c \{km\}}  \B_c^{\{mi\}}.
\label{eq: H0 prime}
\en
It is understood that amplitudes in $\Delta S=-1$ and $\Delta S=0$ transitions are related by the Cabibbo-Kobayashi-Maskawa (CKM) matrix elements, namely
\be
C^{\prime}_i= \frac{V_{cb} V^*_{cd}}{V_{cb} V^*_{cs}} C_i,
\,\,
D^{\prime}_i= \frac{V_{ub} V^*_{ud}}{V_{ub} V^*_{us}} D_i,
\,\,
E^{\prime}_i= \frac{V_{cb} V^*_{cd}}{V_{cb} V^*_{cs}} E_i,
\,\,
F^{\prime}_i= \frac{V_{ub} V^*_{ud}}{V_{ub} V^*_{us}} F_i,
\,\,
A^{\prime}_i= \frac{V_{ub} V^*_{ud}}{V_{ub} V^*_{us}} A_i,
\en
and
it is useful to estimate their relative sizes:
\be
|C^{\prime}_i|\simeq 0.23 |C_i|,
\,\,
|D^{\prime}_i|\simeq  4.33 |D_i|, 
\,\,
|E^{\prime}_i|\simeq 0.23|E_i|, 
\,\,
|F^{\prime}_i|\simeq 4.33 |F_i|, 
\,\,
|A^{\prime}_i|\simeq  4.33 |A_i|.
\en
It is clear that for $b\to u\bar u q$ and $b\to c\bar c q$ ratios, we have
\be
\frac{|A'_i|}{|C'_i|}\simeq 20 \frac{|A_i|}{|C_i|}, \quad
\frac{|D'_i|}{|C'_i|}\simeq 20 \frac{|D_i|}{|C_i|}, \quad
\frac{|F'_i|}{|E'_i|}\simeq 20 \frac{|F_i|}{|E_i|}, 
\label{eq: 20 1}
\en
as noted previously in Eq. (\ref{eq: 20}).

Following ref. \cite{Chua:2026awd}, we can include SU(3) breaking to the above Hamiltonian. 
For the $\Delta S=0$ transition the Hamiltonian with SU(3) breaking is given by
\be
H_{\rm eff}\left(\overline B \to \B_c(\bf {\bar 3_f}) \overline \B_c(\bf { 3_f}) \right)
&=& 
       C_1 \, \overline B_m H^j \overline \B_{c [jk]}  \B_c^{[km]}
       +C_1 \delta^{c_1}_v  \, \overline B_m H^j M_j^{j'}\overline \B_{c [j'k]}  \B_c^{[km]}
 \non\\
 &&+     C_1 \delta^{c_1}_s \, \overline B_{m'} M^{m'}_m H^j \overline \B_{c [jk]}  \B_c^{[km]}
      +   C_1  \delta^{c_1}_c  \, \overline B_m  H^j \overline \B_{c [jk']} M^{k'}_k \B_c^{[km]}
\non\\      
     && +\frac{1}{2} E_1 \, \overline B_j H^j \overline \B_{c [lm]}  \B_c^{[ml]}
     +\frac{1}{2} E_1 \delta^{e_1}_v \, \overline B_{j'} M^{j'}_j H^j \overline \B_{c [lm]}  \B_c^{[ml]}
\non\\
 &&
            +\frac{1}{2}  E_1 \delta^{e_1}_c\, \overline B_j H^j (\overline \B_{c [l'm]}  M^{l'}_l \B_c^{[ml]}+\overline \B_{c [lm']}  M^{m'}_m \B_c^{[ml]})
 \non\\
&&        
            +D_1\,\overline B_m H^{jk}_i \overline \B_{c [jk]}  \B_c^{[mi]}
            +D_1 \delta^{d_1}_v \,\overline B_m H^{jk}_i M^k_{k'} \overline \B_{c [jk]}  \B_c^{[mi]}
\non\\       
&&      
           +D_1 \delta^{d_1}_s \,\overline B_{m'} M^{m'}_m  H^{jk}_i \overline \B_{c [jk]}  \B_c^{[mi]}
           +F_1 \overline B_k H^{jk}_i \overline \B_{c [jm]}  \B_c^{[mi]}
\non\\
&&
         +F_1 \delta^{f_1}_v \overline B_{k'} M^{k'}_k H^{jk}_i \overline \B_{c [jm]}  \B_c^{[mi]}
         +F_1 \delta^{f_1}_c \overline B_{k} H^{jk}_i \overline \B_{c [jm']}  M^{m'}_m \B_c^{[mi]}
\non\\
&&
         +A_1 \overline B_j H^{jk}_i \overline \B_{c [km]}  \B_c^{[mi]} 
        +A_1 \delta^{a_1}_v \overline B_j H^{jk'}_i M^{k}_{k'}\overline \B_{c [km]}  \B_c^{[mi]}
\non\\
&&
       +A_1 \delta^{a_1}_c \overline B_j H^{jk}_i \overline \B_{c [km']} M^{m'}_m \B_c^{[mi]},
\label{eq: H1 I}
 \en
 \be
H_{\rm eff}\left(\overline B \to \B_c(\bf 6_f) \overline \B_c(\bf { 3_f}) \right)
&=& 
       C_2 \, \overline B_m H^j \overline \B_{c\{jk\}}  \B_c^{[km]}       
       + C_2 \delta^{c_2}_v \, \overline B_m H^j M_j^{j'}\overline \B_{c \{j'k\}}  \B_c^{[km]}
\non\\
&&
       + C_2  \delta^{c_2}_s    \, \overline B_{m'} M^{m'}_m H^j \overline \B_{c \{jk\}}  \B_c^{[km]}
       + C_2  \delta^{c_2}_c \, \overline B_m  H^j \overline \B_{c \{jk'\}} M^{k'}_k \B_c^{[km]}
\non\\
&&        
            -D_2\,\overline B_m H^{jk}_i \overline \B_{c \{jk\}}  \B_c^{[mi]}
            -D_2 \delta^{d_2}_v \,\overline B_m H^{jk}_i M^k_{k'} \overline \B_{c \{jk\}}  \B_c^{[mi]}
\non\\       
&&      
           -D_2 \delta^{d_2}_s \,\overline B_{m'} M^{m'}_m  H^{jk}_i \overline \B_{c \{jk\}}  \B_c^{[mi]}
           -F_2 \overline B_k H^{jk}_i \overline \B_{c \{jm\}}  \B_c^{[mi]}
\non\\
&&
         -F_2 \delta^{f_2}_v \overline B_{k'} M^{k'}_k H^{jk}_i \overline \B_{c \{jm\}}  \B_c^{[mi]}
         -F_2 \delta^{f_2}_c \overline B_{k} H^{jk}_i \overline \B_{c \{jm'\}}  M^{m'}_m \B_c^{[mi]}
\non\\
&&
         +A_2 \overline B_j H^{jk}_i \overline \B_{c \{km\}}  \B_c^{[mi]} 
        +A_2 \delta^{a_2}_v \overline B_j H^{jk'}_i M^{k}_{k'}\overline \B_{c \{km'\}}  \B_c^{[mi]}
\non\\
&&
        +A_2 \delta^{a_2}_c \overline B_j H^{jk}_i \overline \B_{c \{km'\}} M^{m'}_m \B_c^{[mi]},
\label{eq: H1 II}        
\en
\be
 H_{\rm eff}\left(\overline B \to \B_c(\bf {\bar 3_f}) \overline \B_c(\bf {\bar 6_f}) \right)
&=& 
       C_3 \, \overline B_m H^j \overline \B_{c[jk]}  \B_c^{\{km\}}
              + C_3  \delta^{c_3}_v\, \overline B_m H^j M_j^{j'}\overline \B_{c [j'k]}  \B_c^{\{km\}}
 \non\\
 &&
      + C_3  \delta^{c_3}_s     \, \overline B_{m'} M^{m'}_m H^j \overline \B_{c [jk]}  \B_c^{\{km\}}
      + C_3 \delta^{c_3}_c     \, \overline B_m  H^j \overline \B_{c [jk']} M^{k'}_k \B_c^{\{km\}}
\non\\
&&        
            -D_3\,\overline B_m H^{jk}_i \overline \B_{c [jk]}  \B_c^{\{mi\}}
            -D_3 \delta^{d_3}_v \,\overline B_m H^{jk}_i M^k_{k'} \overline \B_{c [jk]}  \B_c^{\{mi\}}
\non\\       
&&      
           -D_3 \delta^{d_3}_s \,\overline B_{m'} M^{m'}_m  H^{jk}_i \overline \B_{c [jk]}  \B_c^{\{mi\}}
           -F_3 \overline B_k H^{jk}_i \overline \B_{c [jm]}  \B_c^{\{mi\}}
\non\\
&&
         -F_3 \delta^{f_3}_v \overline B_{k'} M^{k'}_k H^{jk}_i \overline \B_{c [jm]}  \B_c^{\{mi\}}
         -F_3 \delta^{f_3}_c \overline B_{k} H^{jk}_i \overline \B_{c [jm']}  M^{m'}_m \B_c^{\{mi\}}
\non\\
&&
         +A_3 \overline B_j H^{jk}_i \overline \B_{c [km]}  \B_c^{\{mi\}} 
         +A_3 \delta^{a_3}_v \overline B_j H^{jk'}_i M^{k}_{k'}\overline \B_{c [km']}  \B_c^{\{mi\}}
\non\\
&&
        +A_3 \delta^{a_3}_c \overline B_j H^{jk}_i \overline \B_{c [km']} M^{m'}_m \B_c^{\{mi\}},
\label{eq: H1 III}        
\en
\be
 H_{\rm eff}\left(\overline B \to \B_c(\bf 6_f) \overline \B_c(\bf {\bar 6_f}) \right)
&=& 
       C_4 \, \overline B_m H^j \overline \B_{c\{jk\}}  \B_c^{\{km\}}
              + C_4  \delta^{c_4}_v\, \overline B_m H^j M_j^{j'}\overline \B_{c \{j'k\}}  \B_c^{\{km\}}
 \non\\
 &&+  C_4 \delta^{c_4}_s    \, \overline B_{m'} M^{m'}_m H^j \overline \B_{c \{jk\}}  \B_c^{\{km\}}
      + C_4 \delta^{c_4}_c     \, \overline B_m  H^j \overline \B_{c \{jk'\}} M^{k'}_k \B_c^{\{km\}}
\non\\    
   &&   +\frac{1}{2}E_4 \, \overline B_j H^j \overline \B_{c\{lm\}}  \B_c^{\{ml\}}
      +\frac{1}{2}E_4\delta^{e_4}_v  \, \overline B_{j'} M^{j'}_j H^j \overline \B_{c \{lm\}}  \B_c^{\{ml\}}
\non\\
 &&+\frac{1}{2}  E_4 \delta^{e_4}_c\, \overline B_j H^j 
        (\overline \B_{c \{l'm\}}  M^{l'}_l \B_c^{\{ml\}}+\overline \B_{c \{lm'\}}  M^{m'}_m \B_c^{\{ml\}})
\non\\
&&        
            +D_4\,\overline B_m H^{jk}_i \overline \B_{c \{jk\}}  \B_c^{\{mi\}}
            +D_4 \delta^{d_4}_v \,\overline B_m H^{jk}_i M^k_{k'} \overline \B_{c \{jk\}}  \B_c^{\{mi\}}
\non\\       
&&      
           +D_4 \delta^{d_4}_s \,\overline B_{m'} M^{m'}_m  H^{jk}_i \overline \B_{c \{jk\}}  \B_c^{\{mi\}}
           +F_4 \overline B_k H^{jk}_i \overline \B_{c \{jm\}}  \B_c^{\{mi\}}
\non\\
&&
         +F_4 \delta^{f_4}_v \overline B_{k'} M^{k'}_k H^{jk}_i \overline \B_{c \{jm\}}  \B_c^{\{mi\}}
         +F_4 \delta^{f_4}_c \overline B_{k} H^{jk}_i \overline \B_{c \{jm'\}}  M^{m'}_m \B_c^{\{mi\}}
\non\\
&&
         +A_4 \overline B_j H^{jk}_i \overline \B_{c \{km\}}  \B_c^{\{mi\}} 
         +A_4 \delta^{a_4}_v \overline B_j H^{jk'}_i M^{k}_{k'}\overline \B_{c \{km'\}}  \B_c^{\{mi\}}
\non\\
&&
        +A_4 \delta^{a_4}_c \overline B_j H^{jk}_i \overline \B_{c \{km'\}} M^{m'}_m \B_c^{\{mi\}},
\label{eq: H1 IV}
\en
where we have
\begin{equation}
M= \left(
\begin{array}{ccc}
0 &0 &0\\
0 &0 &0 \\
0 &0 &1
\end{array}
\right),
\label{eq: M}
\end{equation}
as the breaking originates from the relatively large $s$-quark mass.
The Hamiltonians for the $\Delta S=0$ transition can be obtained readily.

Note that in the above equations we do not need combinations like $M^{j'}_j H^{jk}_i$ or $M^i_{i'} H^{jk}_i$, 
as one can readily see from Eqs. (\ref{eq: Hjki Hk}) and (\ref{eq: M}), these combinations vanish. 

\begin{figure}[t]
\centering
 \subfigure[]{
  \includegraphics[width=0.45\textwidth]{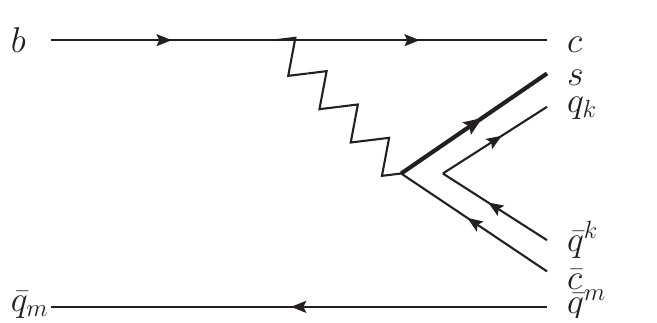}
}
\hspace{12pt}
\subfigure[]{
  \includegraphics[width=0.45\textwidth]{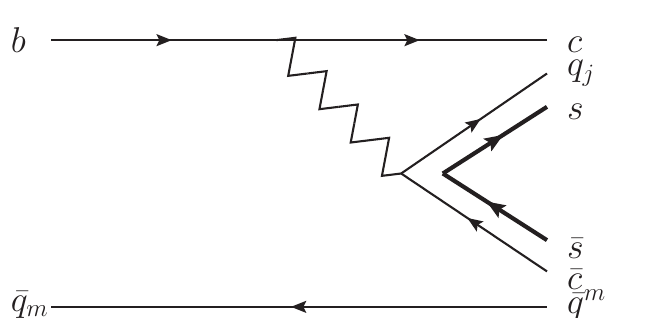}
}
\\\subfigure[]{
  \includegraphics[width=0.45\textwidth]{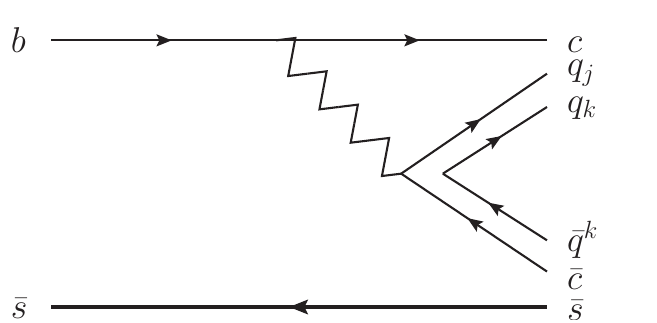}
}
\hspace{12pt}
\subfigure[]{
  \includegraphics[width=0.45\textwidth]{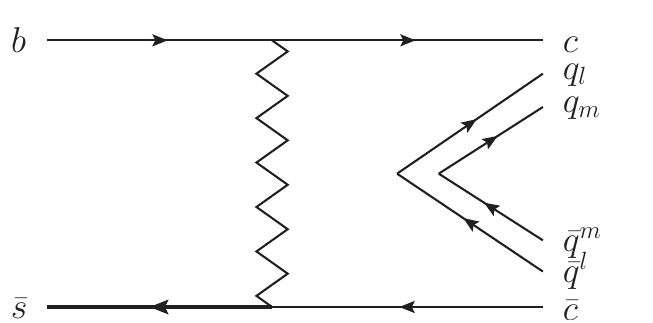}
}
\\\subfigure[]{
  \includegraphics[width=0.45\textwidth]{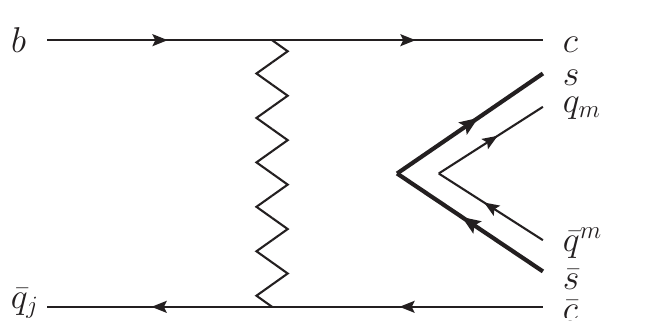}
}
\hspace{12pt}
\subfigure[]{
  \includegraphics[width=0.45\textwidth]{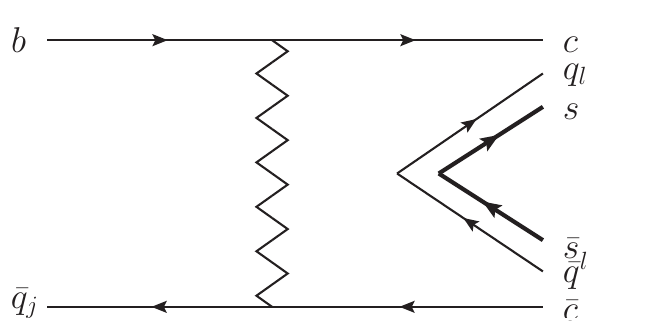}
}
\caption{Internal $W$-tree diagrams containing SU(3) breaking effects,  in
  (a) vertex line, (b) pair creation line, and (c) spectator line;  
  $W$-exchange diagrams containing SU(3) breaking effects, in (d) vertex line and (e)-(f) pair creation lines, in $b\to c\,\bar c q$ transitions.
  These diagrams are the same as those in ref. \cite{Chua:2026awd}.
} \label{fig:TA s}
\end{figure}

\begin{figure}[t]
\centering
 \subfigure[]{
  \includegraphics[width=0.45\textwidth]{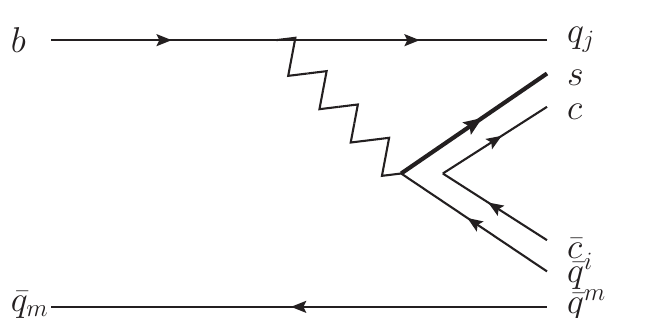}
}
\hspace{12pt}
\subfigure[]{
  \includegraphics[width=0.45\textwidth]{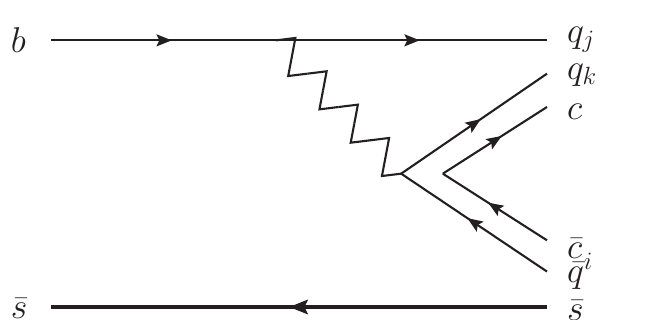}
}
\\
\subfigure[]{
  \includegraphics[width=0.45\textwidth]{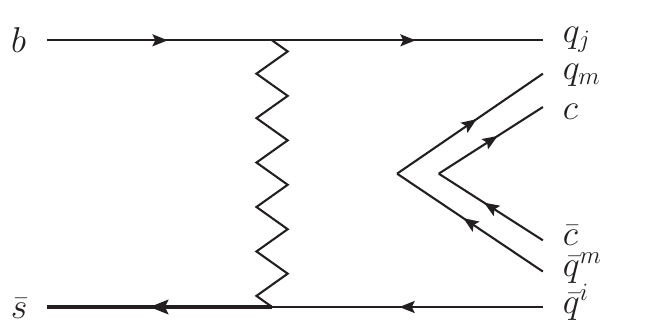}
}
\hspace{12pt}
\subfigure[]{
  \includegraphics[width=0.45\textwidth]{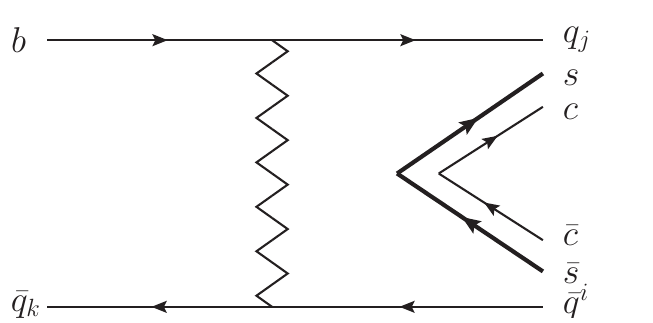}
}
\\
\subfigure[]{
  \includegraphics[width=0.45\textwidth]{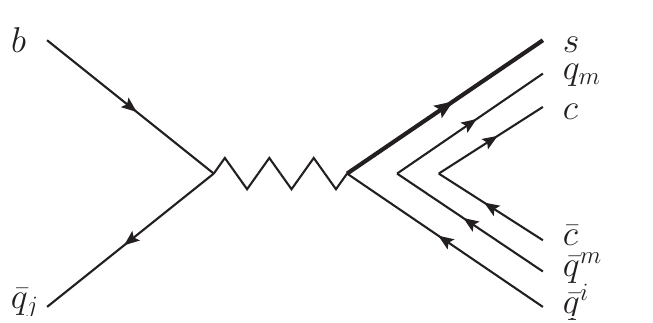}
}
\hspace{12pt}
\subfigure[]{
  \includegraphics[width=0.45\textwidth]{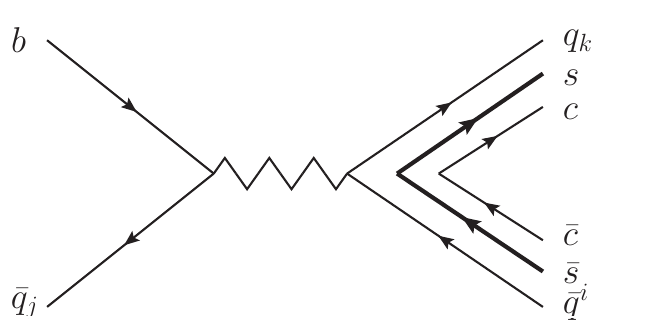}
}
\caption{Internal $W$-tree diagrams containing SU(3) breaking effects,  in
  (a) vertex line and (b) spectator line;  
  $W$-exchange diagrams containing SU(3) breaking effects, in (c) vertex line and (d) pair creation line;
  annihilation diagrams containing SU(3) breaking effects, in (e) vertex line and (f) pair creation line in $b\to u\,\bar u q$ transitions.
  All of the above diagrams are new.
} \label{fig:TA u s}
\end{figure}

To simplify the notation, we define
\be
C^{(\prime)}_{i,v}&\equiv& (1+\delta_v^{c_i}) C^{(\prime)}_i,
\non\\
C^{(\prime)}_{i,s}&\equiv& (1+\delta_s^{c_i}) C^{(\prime)}_i,
\non\\
C^{(\prime)}_{i,c}&\equiv& (1+\delta_c^{c_i}) C^{(\prime)}_i,
\non\\
C^{(\prime)}_{i,vs}&\equiv& (1+\delta_v^{c_i}+\delta_s^{c_i}) C^{(\prime)}_i,
\non\\
C^{(\prime)}_{i,cs}&\equiv& (1+\delta_c^{c_i}+\delta_s^{c_i}) C^{(\prime)}_i,
\non\\
C^{(\prime)}_{i,vcs}&\equiv& (1+\delta_v^{c_i}+\delta_c^{c_i}+\delta_s^{c_i}) C^{(\prime)}_i,
\label{eq: C SU(3) X}
\en
\be
D^{(\prime)}_{i,v}&\equiv& (1+\delta_v^{d_i}) D^{(\prime)}_i,
\non\\
D^{(\prime)}_{i,s}&\equiv& (1+\delta_s^{d_i}) D^{(\prime)}_i,
\non\\
D^{(\prime)}_{i,vs}&\equiv& (1+\delta_v^{d_i}+\delta_s^{d_i}) D^{(\prime)}_i,
\label{eq: D SU(3) X}
\en
\be
F^{(\prime)}_{i,v}&\equiv& (1+\delta_v^{f_i}) F^{(\prime)}_i,
\non\\
F^{(\prime)}_{i,c}&\equiv& (1+\delta_c^{f_i}) F^{(\prime)}_i,
\non\\
F^{(\prime)}_{i,vc}&\equiv& (1+\delta_v^{f_i}+\delta_c^{f_i}) F^{(\prime)}_i,
\label{eq: F SU(3) X}
\en
\be
A^{(\prime)}_{i,v}&\equiv& (1+\delta_v^{a_i}) A^{(\prime)}_i,
\non\\
A^{(\prime)}_{i,c}&\equiv& (1+\delta_c^{a_i}) A^{(\prime)}_i,
\non\\
A^{(\prime)}_{i,vc}&\equiv& (1+\delta_v^{a_i}+\delta_c^{a_i}) A^{(\prime)}_i,
\label{eq: A SU(3) X}
\en
for $i=1-4$, and, likewise,
\be
E^{(\prime)}_{j,v}&\equiv& (1+\delta_v^{e_j}) E^{(\prime)}_{j},
\non\\
E^{(\prime)}_{j,c}&\equiv& (1+\delta_c^{e_j}) E^{(\prime)}_{j},
\non\\
E^{(\prime)}_{j,cc}&\equiv& (1+2\delta_c^{e_j}) E^{(\prime)}_{j},
\non\\
E^{(\prime)}_{j,vc}&\equiv& (1+\delta_v^{e_j}+\delta_c^{e_j}) E^{(\prime)}_j,
\non\\
E^{(\prime)}_{j,vcc}&\equiv& (1+\delta_v^{e_j}+2\delta_c^{e_j}) E^{(\prime)}_j,
\label{eq: E SU(3) X}
\en
for $j=1, 4$.
These $\delta$ parameters correspond to the SU(3) breaking in amplitudes.
In Fig. \ref{fig:TA s}, some topological diagrams in the $b\to c\bar c q$ transition containing SU(3) breaking effects are shown, while
some for the $b\to u\bar u q$ transition are depicted in Fig. \ref{fig:TA u s}.

 \section{Topological amplitudes including $b\to u\bar u q$ contributions}

Using the Hamiltonian in Eqs. (\ref{eq: H1 I}), (\ref{eq: H1 II}), (\ref{eq: H1 III}) and (\ref{eq: H1 IV}), we can obtain the updated decay amplitudes of various $\overline B_q\to \B_c\overline \B_c$ decays.
The updated amplitudes for $\overline B_q\to \B_c({\bf \bar 3_f})\overline \B_c ({\bf 3_f})$ decays,
$\overline B_q\to \B_c({\bf 6_f})\overline \B_c ({\bf 3_f})$ decays,
$\overline B_q\to \B_c({\bf \bar 3_f})\overline \B_c ({\bf \bar 6_f})$ decays
and $\overline B_q\to \B_c({\bf 6_f})\overline \B_c ({\bf \bar 6_f})$ decays 
are given in Tables \ref{tab: BtoBcBcbar I}, \ref{tab: BtoBcBcbar II}, \ref{tab: BtoBcBcbar III} and \ref{tab: BtoBcBcbar IV}, respectively.
Note that although we only show the decays of low-lying anti-triplet and sextet charmed baryons in these tables, 
similar decompositions are applicable for modes with excited-state anti-triplet and sextet charmed baryons. 
These tables are some of the main results of this work.

Note that, as shown in Tables \ref{tab: BtoBcBcbar II} and \ref{tab: BtoBcBcbar III}, we have four new modes in $\overline B_q\to \B_c({\bf 6_f})\overline \B_c ({\bf 3_f})$ decays
and $\overline B_q\to \B_c({\bf \bar 3_f})\overline \B_c ({\bf \bar 6_f})$ decays. 
They are $\overline B{}_s^0\to \Sigma_c^+ \bar \Lambda_c^{-}$ and $\overline B{}^0\to \Xi_c^{\prime +} \bar \Xi_c^{-}$ decays in Table \ref{tab: BtoBcBcbar II},
and $\overline B{}_s^0\to \Lambda_c^+ \bar \Sigma_c^{-}$ and $\overline B{}^0\to \Xi_c^+ \bar \Xi_c^{\prime -}$ decays in Table \ref{tab: BtoBcBcbar III}.
They only have exchange diagrams from $b\to u\bar u q$ contributions.

\begin{table}[t!]
\caption{\label{tab: BtoBcBcbar I}
Updated
$\overline B_q\to \B_c({\bf \bar 3_f})\overline \B_c ({\bf 3_f})$ decay amplitudes in $\Delta S=-1$ and  $\Delta S=0$ transitions.}
\footnotesize{
\begin{ruledtabular}
\begin{tabular}{llcllccccr}
Mode
          & $A \left(\overline B_q\to \B_c({\bf \bar 3_f})\overline \B_c ({\bf 3_f})\right)$
          & Mode
          & $A \left(\overline B_q\to \B_c({\bf \bar 3_f})\overline \B_c ({\bf 3_f})\right)$
          \\
\hline
$B^-\to \Xi_c^0 \bar \Lambda_c^{-}$
          & $C_{1,v}+A_{1,v}$     
          & $\overline B{}^0\to \Xi_c^+ \bar \Lambda_c^{-}$
          & $C_{1,v}+D_{1,v}$ 
          \\
$\overline B{}_s^0\to \Xi_c^0 \bar \Xi_c^{0}$
          & $-C_{1,vs}-E_{1,vc}$ 
          & $\overline B{}_s^0\to \Xi_c^+ \bar \Xi_c^{-}$
          & $-C_{1,vs}-D_{1,vs}-E_{1,vc}-F_{1,vc}$ 
          \\
$\overline B{}_s^0\to \Lambda_c^+ \bar \Lambda_c^{-}$
          & $-E_{1,v}-F_{1,v}$ 
          \\
\hline
$B^-\to \Xi_c^0 \bar \Xi_c^{-}$
          & $C'_{1,c}+A'_{1,c}$ 
          & $\overline B{}^0\to \Lambda_c^+ \bar \Lambda_c^{-}$
          & $-C'_1-D'_1-E'_1-F'_1$ 
          \\          
          $\overline B{}^0\to \Xi_c^0 \bar \Xi_c^{0}$
          & $-C'_{1,c}-E'_{1,c}$  
          & $\overline B{}^0\to \Xi_c^+ \bar \Xi_c^{-}$
          & $-E'_{1,c}-F'_{1,c}$          
          \\ 
          $\overline B{}_s^0\to \Lambda_c^+ \bar \Xi_c^{-}$
          & $C'_{1,s}+D'_{1,s}$ 
          \\                                            
\end{tabular}
\end{ruledtabular}
}
\end{table}

\begin{table}[t!]
\caption{\label{tab: BtoBcBcbar II}
Updated
$\overline B_q\to \B_c({\bf 6_f})\overline \B_c ({\bf 3_f})$ decay amplitudes in $\Delta S=-1$ and  $\Delta S=0$ transitions.}
\footnotesize{
\begin{ruledtabular}
\begin{tabular}{lclc}
Mode
          & $A \left(\overline B_q\to \B_c({\bf 6_f})\overline \B_c ({\bf 3_f})\right)$
          & Mode
          & $A \left(\overline B_q\to \B_c({\bf 6_f})\overline \B_c ({\bf 3_f})\right)$
          \\
\hline
           $B^-\to \Xi_c^{\prime 0} \bar \Lambda_c^{-}$
          & $-C_{2,v}-A_{2,v}$ 
          & $B^-\to \Omega_c^{0} \bar \Xi_c^{-}$
          & $\sqrt2 (C_{2,vc}+A_{2,vc})$ 
          \\     
          $\overline B{}^0\to \Xi_c^{\prime +} \bar \Lambda_c^{-}$
          & $C_{2,v}+D_{2,v}$
          & $\overline B{}^0\to \Omega_c^{0} \bar \Xi_c^{0}$
          & $-\sqrt2 C_{2,vc}$
          \\
          $\overline B{}_s^0\to \Xi_c^{\prime +} \bar \Xi_c^{-}$
          & $-C_{2,vs}-D_{2,vs}-F_{2,vc}$
          & $\overline B{}_s^0\to \Xi_c^{\prime 0} \bar \Xi_c^{0}$
          & $C_{2,vs}$
           \\
           $\overline B{}_s^0\to \Sigma_c^+ \bar \Lambda_c^{-}$
          & $F_{2,v}$
           \\
\hline
         $B^-\to \Sigma_c^0 \bar \Lambda_c^{-}$
          & $-\sqrt2 (C'_2+A'_2)$
          & $B^-\to \Xi_c^{\prime 0} \bar \Xi_c^{-}$
          & $C'_{2,c}+A'_{2,c}$
         \\
         $\overline B{}^0\to \Sigma_c^+ \bar \Lambda_c^{-}$
          & $C'_2+D'_2+F'_2$    
          & $\overline B{}^0\to \Xi_c^{\prime 0} \bar \Xi_c^{0}$
          & $-C'_{2,c}$
          \\ 
$\overline B{}^0\to \Xi_c^{\prime +} \bar \Xi_c^{-}$
          & $-F'_{2,c}$ 
          & $\overline B{}_s^0\to \Sigma_c^+ \bar \Xi_c^{-}$
          & $-C'_{2,s}-D'_{2,s}$
          \\   
$\overline B{}_s^0\to \Sigma_c^0 \bar \Xi_c^{0}$
          & $\sqrt2 C'_{2,s}$                        
\end{tabular}
\end{ruledtabular}
}
\end{table}

\begin{table}[t!]
\caption{\label{tab: BtoBcBcbar III}
Updated
$\overline B_q\to \B_c({\bf \bar 3_f})\overline \B_c ({\bf \bar 6_f})$ decay amplitudes in $\Delta S=-1$ and  $\Delta S=0$ transitions.}
\footnotesize{
\begin{ruledtabular}
\begin{tabular}{lclc}
Mode
          & $A \left(\overline B_q\to \B_c({\bf \bar 3_f})\overline \B_c ({\bf \bar 6_f})\right)$
          & Mode
          & $A \left(\overline B_q\to \B_c({\bf \bar 3_f})\overline \B_c ({\bf \bar 6_f})\right)$
          \\
\hline
          $B^-\to \Xi_c^+ \bar \Sigma_c^{--}$
          & $\sqrt2 (C_{3,v}+D_{3,v}+A_{3,v})$ 
          & $B^-\to \Xi_c^0 \bar \Sigma_c^{-}$
          & $-C_{3,v}-A_{3,v}$ 
          \\     
          $\overline B{}^0\to \Xi_c^0 \bar \Sigma_c^{0}$
          & $-\sqrt2 C_{3,v}$
          & $\overline B{}^0\to \Xi_c^+ \bar \Sigma_c^{-}$
          & $C_{3,v}+D_{3,v}$
          \\
          $\overline B{}_s^0\to \Xi_c^0 \bar \Xi_c^{\prime 0}$
          & $-C_{3,vs}$
          & $\overline B{}_s^0\to \Xi_c^+ \bar \Xi_c^{\prime -}$
          & $C_{3,vs}+D_{3,vs}+F_{3,vc}$
          \\
          $\overline B{}_s^0\to \Lambda_c^+ \bar \Sigma_c^{-}$
          & $-F_{3,v}$
          \\
\hline
         $B^-\to \Lambda_c^+ \bar \Sigma_c^{- -}$
          & $-\sqrt2 (C'_3+D'_3+A'_3)$
          & $B^-\to \Xi_c^0 \bar \Xi_c^{\prime -}$
          & $C'_{3,c}+A'_{3,c}$
          \\
          $\overline B{}^0\to \Lambda_c^+ \bar \Sigma_c^{-}$
          & $-(C'_3+D'_3+F'_3)$
          & $\overline B{}^0_s\to \Lambda_c^+ \bar \Xi_c^{\prime -}$
          & $-C'_{3,s}-D'_{3,s}$           \\
$\overline B{}^0\to \Xi_c^0 \bar \Xi_c^{\prime 0}$
          & $C'_{3,c}$        
          & $\overline B{}^0\to \Xi_c^+ \bar \Xi_c^{\prime -}$
          & $F'_{3,c}$ 
          \\        
$\overline B{}_s^0\to \Xi_c^0 \bar \Omega_c^{0}$
          & $\sqrt2 C'_{3,cs}$
          \\                                            
\end{tabular}
\end{ruledtabular}
}
\end{table}

\begin{table}[t!]
\caption{\label{tab: BtoBcBcbar IV}
Updated
$\overline B_q\to \B_c({\bf 6_f})\overline \B_c ({\bf \bar 6_f})$ decay amplitudes in $\Delta S=-1$ and  $\Delta S=0$ transitions.}
\footnotesize{
\begin{ruledtabular}
\begin{tabular}{lclc}
Mode
          & $A \left(\overline B_q\to \B_c({\bf 6_f})\overline \B_c ({\bf \bar 6_f})\right)$
          & Mode
          & $A \left(\overline B_q\to \B_c({\bf 6_f})\overline \B_c ({\bf \bar 6_f})\right)$
          \\
\hline
           $B^-\to \Xi_c^{\prime +} \bar \Sigma_c^{--}$
          & $\sqrt2 (C_{4,v}+D_{4,v}+A_{4,v})$ 
          & $B^-\to \Xi_c^{\prime 0} \bar \Sigma_c^{-}$
          & $C_{4,v}+A_{4,v}$ 
          \\     
          $B^-\to \Omega_c^{0} \bar \Xi_c^{\prime -}$
          & $\sqrt2 (C_{4,vc}+A_{4,vc})$ 
          & $\overline B{}^0\to \Xi_c^{\prime +} \bar \Sigma_c^{-}$
          & $C_{4,v}+D_{4,v}$
          \\
          $\overline B{}^0\to \Xi_c^{\prime 0} \bar \Sigma_c^{0}$
          & $\sqrt2 C_{4,v}$
          & $\overline B{}^0\to \Omega_c^{0} \bar \Xi_c^{\prime 0}$
          & $\sqrt2 C_{4,vc}$
          \\
          $\overline B{}_s^0\to \Sigma_c^{++} \bar \Sigma_c^{--}$
          & $E_{4,v}+2 F_{4,v}$
          & $\overline B{}_s^0\to \Sigma_c^{+} \bar \Sigma_c^{-}$
          & $E_{4,v}+F_{4,v}$
          \\
         $\overline B{}_s^0\to \Sigma_c^{0} \bar \Sigma_c^{0}$
          & $E_{4,v}$
          & $\overline B{}_s^0\to \Xi_c^{\prime +} \bar \Xi_c^{\prime -}$
          & $C_{4,vs}+D_{4,vs}+E_{4,vc}+F_{4,vc}$
          \\
          $\overline B{}_s^0\to \Xi_c^{\prime 0} \bar \Xi_c^{\prime 0}$
          & $C_{4,vs}+E_{4,vc}$
          & $\overline B{}_s^0\to \Omega_c^{0} \bar \Omega_c^{0}$
          & $2C_{4,vcs}+E_{4,vcc}$
          \\
\hline
         $B^-\to \Sigma_c^+ \bar \Sigma_c^{- -}$
          & $\sqrt2 (C'_4+D'_4+A'_4)$
          & $B^-\to \Sigma_c^0 \bar \Sigma_c^{ -}$
          & $\sqrt2 (C'_4+A'_4)$
          \\
          $B^-\to \Xi_c^{\prime 0} \bar \Xi_c^{\prime -}$
          & $C'_{4,c}+A'_{4,c}$
          & $\overline B{}^0\to \Sigma_c^{++} \bar \Sigma_c^{ --}$
          & $E'_4+2F'_4$
          \\
          $\overline B{}^0\to \Sigma_c^+ \bar \Sigma_c^{-}$
          & $C'_4+D'_4+E'_4+F'_4$
          & $\overline B{}^0\to \Sigma_c^0 \bar \Sigma_c^{0}$
          & $2 C'_4+E'_4$
          \\
          $\overline B{}^0\to \Xi_c^{\prime +} \bar \Xi_c^{\prime -}$
          & $E'_{4,c}+F'_{4,c}$ 
          & $\overline B{}^0\to \Xi_c^{\prime 0} \bar \Xi_c^{\prime 0}$
          & $C'_{4,c}+E'_{4,c}$ 
          \\          
          $\overline B{}^0\to \Omega_c^0 \bar \Omega_c^{0}$
          & $E'_{4,cc}$  
          & $\overline B{}_s^0\to \Sigma_c^+ \bar \Xi_c^{\prime -}$
          & $C'_{4,s}+D'_{4,s}$
          \\
          $\overline B{}_s^0\to \Sigma_c^0 \bar \Xi_c^{\prime 0}$
          & $\sqrt2 C'_{4,s}$
          & $\overline B{}_s^0\to \Xi_c^{\prime 0} \bar \Omega_c^{0}$
          & $\sqrt2 C'_{4,cs}$
          \\                                              
\end{tabular}
\end{ruledtabular}
}
\end{table}

It will be useful to express these topological amplitudes more explicitly.
For example, we have the following expressions for some of the amplitudes $A\left(\overline B_q\to \B_c({\bf \bar 3_f})\overline \B_c ({\bf 3_f})\right)$ decays shown in Table \ref{tab: BtoBcBcbar I},
\be
C_{1,v}= \frac{G_F}{\sqrt2} m_{B_q} V_{cb} V^*_{cs} c_{1,v},
\quad
C'_{1,v}
=\frac{G_F}{\sqrt2} m_{B_q} V_{cb} V^*_{cd}  c_{1,v},
\non\\
E_{1,v}= \frac{G_F}{\sqrt2} m_{B_q} V_{cb} V^*_{cs} e_{1,v},
\quad
E'_{1,v}
=\frac{G_F}{\sqrt2} m_{B_q} V_{cb} V^*_{cd}  e_{1,v},
\label{eq: b c}
\en
for $b\to c \bar c s (d)$ transition,
and
\be
D_{1,v}= \frac{G_F}{\sqrt2} m_{B_q} V_{ub} V^*_{us} d_{1,v},
\quad
D'_{1,v}
=\frac{G_F}{\sqrt2} m_{B_q} V_{ub} V^*_{ud}  d_{1,v},
\non\\
F_{1,v}= \frac{G_F}{\sqrt2} m_{B_q} V_{ub} V^*_{us} f_{1,v},
\quad
F'_{1,v}
=\frac{G_F}{\sqrt2} m_{B_q} V_{ub} V^*_{ud}  f_{1,v},
\non\\
A_{1,v}= \frac{G_F}{\sqrt2} m_{B_q} V_{ub} V^*_{us} a_{1,v},
\quad
A'_{1,v}
=\frac{G_F}{\sqrt2} m_{B_q} V_{ub} V^*_{ud}  a_{1,v},
\label{eq: b a}
\en
for $b\to u \bar u s (d)$ transition,
with $c_{1,v}=(1+\delta^{c_1}_v) c_{1}$, 
$e_{1,v}=(1+\delta^{e_1}_v) e_{1}$, 
$d_{1,v}=(1+\delta^{d_1}_v) d_{1}$, 
$f_{1,v}=(1+\delta^{f_1}_v) f_{1}$, 
$a_{1,v}=(1+\delta^{a_1}_v) a_{1}$, where the subscripts $v$ denote that SU(3) breaking effect from the weak vertex are included, 
see Fig. \ref{fig:TA s} (a), (d)  and Fig. \ref{fig:TA u s} (a), (c), (e).

\section{Numerical results}


Numerical results on rates of $\overline B_q\to \B_c({\bf \bar 3_f})\overline \B_c ({\bf 3_f})$ decays for low-lying charmed baryons 
will be presented in this section.
Masses and lifetimes of $B_q$ mesons and masses of $\B_c$ baryons and CKM matrix elements are taken from the Particle Data Group, ref. \cite{ParticleDataGroup:2024cfk},
and the latest fit in the CKMfitter Group, ref. \cite{CKMfitter}, respectively.
Some comments on other modes will be given at the end of this section.

\subsection{$\overline B_q\to \B_c({\bf \bar 3_f})\overline \B_c ({\bf 3_f})$ decay rates}

Experimental results of all available $\overline B_q\to \B_c({\bf \bar 3_f})\overline \B_c ({\bf 3_f})$ decay rates for low-lying charmed baryons are used in the fit.
Presently, only data on the following four such decay rates, $B^-\to \Xi_c^0 \bar \Lambda_c^{-}$,
$\overline B{}^0\to \Xi_c^+ \bar \Lambda_c^{-}$,
$\overline B{}_s^0\to \Lambda_c^+ \bar \Lambda_c^{-}$
and
$\overline B{}^0\to \Lambda_c^+ \bar \Lambda_c^{-}$
are reported.
To reduce the number of free parameters, following ref. \cite{Chua:2026awd}, 
we assume $\delta^{e_1}_v=-\delta^{c_1}_v$ for these SU(3) breaking parameters. 
Furthermore, as $D_1$ and $C_1$ are internal-$W$ diagrams, while $F_1$ and $E_1$ are $W$-exchange diagrams, see Fig. (\ref{fig:TA}),
we assume $\delta^{d_1}_v=\delta^{c_1}_v$ and $\delta^{f_1}_v=\delta^{e_1}_v$, as it is reasonable to expect them to have similar SU(3) breaking behavior.
These relations can be relaxed when more data are available.

The number of fitted parameters is still larger than the number of available data in $\overline B_q\to \B_c({\bf \bar 3_f})\overline \B_c ({\bf 3_f})$ decay rates.  
As this work concentrates on the $b\to u\bar u q$ contributions, 
we take parameters related to $b\to c\bar c q$ contributions as input
and take the $b\to u\bar u q$ parameters as fitted parameters.
We consider two different cases.
In the first case, we use the central values of the fitted parameters of $b\to c \bar c q$ contributions obtained in ref. \cite{Chua:2026awd} as input parameters, namely the central values of $c_1$, $e_1$ and $\delta^{c_1}_v$.
The three parameters left, namely $d_1$, $f_1$ and $a_{1,v}$, are fitted parameters.
As we have four data points and three fitted parameters, we can perform a chi-square fit.
The uncertainties in the fitted parameters are obtained by scanning the parameter space around the best-fit values with $\chi^2\leq \chi^2_{\rm min.}+1$
and they will give related uncertainties on rates.
We take the uncertainties of other parameters, such as other SU(3) breaking parameters unconstrained by the fit, to be at most $|\delta^{c_1}_v|$, as in ref.~\cite{Chua:2026awd}
and the uncertainties of $c_1$, $e_1$ and $\delta^{c_1}_v$ reported in ref. \cite{Chua:2026awd}.
These uncertainties will propagate to further uncertainties on rates.
In Table \ref{tab: c1}, we show the fitted parameters and $\chi^2_{\rm min.}$
We obtain $\chi^2_{\rm min.}=0.198$ in the first case.
We find that it is possible to obtain $\chi^2_{\rm min.}=0.000$ by slightly modify the input parameters, $c_1$, $e_1$ and $\delta^{c_1}_v$,
and this corresponds to case 2. The uncertainties of these parameters can be taken to be of similar sizes to those in case 1.
We show these two cases in Table~\ref{tab: c1}.

\begin{table}[t!]
\caption{\label{tab: c1}
Parameters in $\overline B_q\to \B_c({\bf \bar 3_f})\overline \B_c ({\bf 3_f})$ for low lying $\B_c({\bf \bar 3_f})$ and $\overline \B_c ({\bf 3_f})$,
where $b\to u\bar u q$ parameters, namely, $f_1$, $d_1$ and $a_{1,v}$, are fitted parameters. 
Two cases are shown.
For the first case the central values of the fitted $b\to c\bar c q$ parameters obtained in ref. \cite{Chua:2026awd} are used as input,
while these input parameters are slightly modified in case 2.
Input parameters are marked by asterisks.
The uncertainties in fitted parameters are obtained by scanning the parameter space around the best-fit values with $\chi^2\leq \chi^2_{\rm min.}+1$.
Parameters marked with daggers provide further uncertainties in rates.
For example, we assume other SU(3) breaking parameters unconstrained by the fit have sizes at most as $|\delta^{c_1}_v|$.
See text for details.
}
\begin{ruledtabular}
\begin{tabular}{ccccccccc}         
&$c_1$
          & $e_1$
          & $\delta^{c_1}_v$
          & $d_1$
          & $f_1$
          & $a_{1,v}$
          \\
\hline 
case 1
         & $0.209^*{}_{-0.016}^{+0.015}{}^\dagger$
         & $-0.091^*{}\pm 0.015{}^\dagger$
         & $0.35^*{}\pm 0.11{}^\dagger$
         & $0.450_{-1.409}^{+1.127}$
         & $-0.204_{-1.124}^{+1.405}$ 
         & $-1.009_{-4.442}^{+3.257}$
          \\           
case 2
         &$0.212^*\pm 0.015^\dagger$
         & $-0.083^*\pm0.015^\dagger$
         & $0.39^* \pm 0.11^\dagger$
         & $1.422_{-1.110}^{+0.925}$
         & $-1.312_{-0.963}^{+1.156}$ 
         & $-3.518_{-8.207}^{+4.528}$
          \\     
\hline                   
         &$\delta^{e_1}_v(=-\delta^{c_1}_v)$ 
          &$\delta^{d_1}_v(=\delta^{c_1}_v)$ 
          & $\delta^{f_1}_v(=\delta^{e_1}_v)$
          & $\delta^{c_1}_c$
          & $\delta^{c_1}_s$ 
          & $\delta^{e_1}_c$
\\  
 \hline  
case 1
          &$(-0.35^*{}\pm 0.11{}^\dagger)$
          & $(0.35^*{}\pm 0.11{}^\dagger)$
          & $(-0.35^*{}\pm 0.11{}^\dagger)$
          & $0\pm 0.35^\dagger$
          & $0\pm 0.35^\dagger$
          & $0\pm 0.35^\dagger$
           \\    
case 2
          &$(-0.39^*{}\pm 0.11{}^\dagger)$
          & $(0.39^*{}\pm 0.11{}^\dagger)$
          & $(-0.39^*{}\pm 0.11{}^\dagger)$
          & $0\pm 0.40^\dagger$
          & $0\pm 0.40^\dagger$
          & $0\pm 0.40^\dagger$
           \\   
\hline
          & $\delta^{d_1}_c$
          & $\delta^{d_1}_s$ 
          & $\delta^{f_1}_c$
          & $\delta^{a_1}_v$
          & $\delta^{a_1}_s$
          & $\chi^2_{\rm min.}$
         \\ 
\hline
case 1
          & $0\pm0.35^\dagger$
          & $0\pm 0.35^\dagger$
          & $0\pm 0.35^\dagger$
          & $0\pm 0.35^\dagger$
          & $0\pm 0.35^\dagger$ 
          & 0.198   
          \\
case 2
          & $0\pm0.40^\dagger$
          & $0\pm 0.40^\dagger$
          & $0\pm 0.40^\dagger$
          & $0\pm 0.40^\dagger$
          & $0\pm 0.40^\dagger$ 
          & 0.000                    
\end{tabular}
\end{ruledtabular}
\end{table}

Note that in case 1 the values of $d_1$, $f_1$ and $a_{1,v}$ can vanish.
Hence the $b\to u\bar u q$ contributions can vanish as well, and the results in ref. \cite{Chua:2026awd} can be included in this case.
This is reasonable, as we use the central values of $b\to c\bar c q$ parameters in ref. \cite{Chua:2026awd} as input, which has $\chi^2_{\rm min.}=0.315$.
It is within the range of  $\chi^2\leq \chi^2_{\rm min.}+1$ of case 1.

It is interesting that by slightly modifying the input values of the $b\to c\bar c q$ parameters, we can have a perfect fit to data in case 2.
In this case, we see that $d_1$ and $f_1$ can no longer have vanishing values, while $a_{1,v}$ can still vanish.
The $b\to u\bar u q$ contributions cannot all vanish in this case.
Hence, the present data slightly prefer case 2 over case 1.
In other words, the case with non-vanishing $b\to u\bar u q$ contributions is slightly more preferred.

In both cases the uncertainties in $d_1$, $f_1$ and $a_{1,v}$ are large.
This is understandable as three of the four data are in the $\Delta S=-1$ sector, where the rates are not sensitive to these $b\to u\bar u q$ contributions.
We also note that the signs of $d_1$ and $f_1$ are opposite in both cases. 
Hence, internal-$W$ and $W$-exchange contributions have destructive interference.
This is similar to the situation of $c_1$ and $e_1$, which also correspond to internal-$W$ and $W$-exchange contributions, respectively.
Consequently, the amplitudes containing $D^{(\prime)}_1+F^{(\prime)}_1$ should have cancelation.
These modes are $\overline B{}_s^0\to \Xi_c^+ \bar \Xi_c^{-}$ and $\overline B{}^0\to \Lambda_c^+ \bar \Lambda_c^{-}$ decays, see Table~\ref{tab: BtoBcBcbar I}.
In fact, it is the data in the $\Delta S=-1$ sector, namely the $\overline B{}^0\to \Lambda_c^+ \bar \Lambda_c^{-}$ decay rate, that leads to such cancelation.
We will return to this later.

The sizes of $d_1$, $f_1$ and $a_{1,v}$ are generally larger than those in $c_1$ and $e_1$, especially in case~2.
We do not have an explanation for the physical origin.
However, we note that in the $b\to u\bar u q$ transition diagrams the $c\bar c$ pairs are always pair-created, see Fig.~\ref{fig:TA} (c)-(e).
Perhaps a $c\bar c$ creation mechanism that is proportional to $m_c$ might be responsible for the enhancement.

\begin{table}[t!]
\caption{\label{tab: ratios}
Ratios of the sizes of $b\to u\bar u q$ and $b\to c\bar c q$ topological amplitudes in $\Delta S=-1$ and $\Delta S=0$ transitions in case 1 and case 2.
}
\begin{ruledtabular}
\begin{tabular}{cccc}  
$\Delta S=-1$
        & $\left| A_1 / C_1  \right|$     
        & $\left| D_1 / C_1 \right|$
        & $\left| F_1 / E_1 \right|$
          \\    
\hline
 case 1
        & $0.100_{-0.100}^{+0.441}{}_{-0.031}^{+0.066}$      
        & $0.045_{-0.045}^{+0.112}{}_{-0.003}^{+0.004}$ 
        & $0.047_{-0.047}^{+0.257}{}_{-0.007}^{+0.009}$ 
          \\         
case 2
        & $0.344_{-0.344}^{+0.803}{}_{-0.114}^{+0.266}$      
        & $0.139_{-0.109}^{+0.091}{}_{-0.009}^{+0.011}$ 
        & $0.328_{-0.289}^{+0.241}{}_{-0.050}^{+0.072}$ 
          \\   \hline 
$\Delta S=0$
        & $\left| A'_1 / C'_1  \right|$     
        & $\left| D'_1 / C'_1 \right|$ 
        & $\left| F'_1 / E'_1 \right|$
        \\
\hline
case 1
        & $1.879_{-1.879}^{+8.273}{}_{-0.579}^{+1.246}$ 
        & $0.839_{-0.839}^{+2.099}{}_{-0.056}^{+0.070}$ 
        & $0.875_{-0.875}^{+4.829}{}_{-0.124}^{+0.174}$ 
         \\                                          
case 2
        & $6.459_{-6.459}^{+15.070}{}_{-2.129}^{+4.996}$ 
        & $2.611_{-2.038}^{+1.698}{}_{-0.173}^{+0.199}$ 
        & $6.155_{-5.425}^{+4.521}{}_{-0.943}^{+1.359}$ 
         \\      
         \end{tabular}
\end{ruledtabular}
\end{table}

\begin{table}[t!]
\caption{\label{tab: Br BtoBcBcbar 3bar3}
$\overline B_q\to \B_c({\bf \bar 3_f})\overline \B_c ({\bf 3_f})$ branching ratios (in unit of $10^{-4}$) in $\Delta S=-1$ and  $\Delta S=0$ transitions in case 1 and case 2.
Experimental results are inputs to the $\chi^2$-fit. 
The first uncertainties are from the uncertainties of the fitted parameters, namely $d_1$, $e_1$ and $a_{1,v}$, 
by scanning the parameter space around the best-fit values with $\chi^2\leq \chi^2_{\rm min.}+1$,
while the second uncertainties are from other parameters, see Table~\ref{tab: c1}.
Decay amplitudes from Table~\ref{tab: BtoBcBcbar I} are shown for convenience.
Results from ref. \cite{Chua:2026awd} are also shown for comparison.
}
\footnotesize{
\begin{ruledtabular}
\begin{tabular}{cccccc}
Mode
          & $A \left(\overline B_q\to \B_c({\bf \bar 3_f})\overline \B_c ({\bf 3_f})\right)$
          & $Br$
          & $Br$
          & $Br$
          & Expt.
           \\
          & 
          & case 1
          & case 2
          & ref. \cite{Chua:2026awd}
          &
           \\
\hline
$B^-\to \Xi_c^0 \bar \Lambda_c^{-}$
          & $C_{1,v}+A_{1,v}$  
          & $9.51_{-1.09}^{+2.17}{}_{-2.73}^{+3.36}$
          & $9.51_{-0.27}^{+2.22}{}_{-2.64}^{+3.35}$
          & $10.06_{-2.04}^{+1.98}$
          & $9.51\pm 2.28$~\cite{Belle:2018kzz}
          \\
$\overline B{}^0\to \Xi_c^+ \bar \Lambda_c^{-}$
          & $C_{1,v}+D_{1,v}$
          & $9.70_{-0.99}^{+1.05}{}_{-2.70}^{+3.30}$
          & $11.60\pm1.09_{-3.02}^{+3.75}$
          & $9.34_{-1.90}^{+1.84}$
          & $11.6\pm 4.46$~\cite{Belle:2019bgi}
          \\
$\overline B{}_s^0\to \Lambda_c^+ \bar \Lambda_c^{-}$
          & $-E_{1,v}-F_{1,v}$
          & $0.52_{-0.09}^{+0.15}{}_{-0.27}^{+0.44}$
          & $0.50_{-0.12}^{+0.15}{}_{-0.26}^{+0.42}$
          & $0.50_{-0.15}^{+0.16}$
          & $0.50\pm 0.16$~ \cite{LHCb:2025ueu}
          \\
$\overline B{}_s^0\to \Xi_c^+ \bar \Xi_c^{-}$
          & $-C_{1,vs}-D_{1,vs}-E_{1,vc}-F_{1,vc}$
          & $5.56_{-0.40}^{+0.39}{}_{-5.22}^{+11.46}$
          & $7.07_{-0.43}^{+0.41}{}_{-6.61}^{+14.54}$ 
          & $5.36_{-4.46}^{+8.72}$
          & $-$
          \\
$\overline B{}_s^0\to \Xi_c^0 \bar \Xi_c^{0}$
          & $-C_{1,vs}-E_{1,vc}$
          & $5.33\pm 0.00_{-5.01}^{+11.02}$
          & $6.44\pm0.00_{-5.96}^{+12.79}$ 
          & $5.33_{-4.44}^{+8.67}$
          & $-$
          \\     
           \hline
$\overline B{}^0\to \Lambda_c^+ \bar \Lambda_c^{-}$
          & $-C'_1-D'_1-E'_1-F'_1$
          & $0.101_{-0.017}^{+0.032}{}_{-0.029}^{+0.041}$
          & $0.101_{-0.001}^{+0.031}{}_{-0.042}^{+0.055}$
          & $0.101_{-0.032}^{+0.031}$
          & $0.101\pm0.032$~\cite{LHCb:2025ueu}
          \\          
$B^-\to \Xi_c^0 \bar \Xi_c^{-}$
          & $C'_{1,c}+A'_{1,c}$
          & $1.45_{-1.25}^{+25.53}{}_{-1.08}^{+4.52}$
          & $11.86_{-11.65}^{+107.34}{}_{-9.52}^{+48.80}$
          & $0.240_{-0.154}^{+0.261}$
          & $-$
          \\
$\overline B{}_s^0\to \Lambda_c^+ \bar \Xi_c^{-}$
          & $C'_{1,s}+D'_{1,s}$
          & $0.304_{-0.056}^{+1.849}{}_{-0.187}^{+0.304}$
          & $1.74_{-1.48}^{+3.17}{}_{-1.10}^{+1.89}$
          & $0.297_{-0.190}^{+0.322}$
          & $-$
          \\                                            
$\overline B{}^0\to \Xi_c^+ \bar \Xi_c^{-}$
          & $-E'_{1,c}-F'_{1,c}$  
          & $0.045_{-0.009}^{+1.171}{}_{-0.029}^{+0.055}$
          & $1.19_{-1.11}^{+2.55}{}_{-0.78}^{+1.28}$ 
          & $0.042_{-0.030}^{+0.063}$ 
          & $-$
          \\  
$\overline B{}^0\to \Xi_c^0 \bar \Xi_c^{0}$
          & $-C'_{1,c}-E'_{1,c}$  
          & $0.071\pm0.000_{-0.071}^{+0.254}$
          & $0.085\pm0.000_{-0.085}^{+0.300}$
          & $0.071_{-0.071}^{+0.227}$
          & $-$
          \\
\hline          
$\chi^2_{\rm min.}$
        &
        & $0.198$
        & $0.000$
        & $0.315$
\end{tabular}
\end{ruledtabular}
}
\end{table}

We show in Table~\ref{tab: ratios}
the ratios of the sizes of $b\to u\bar u q$ and $b\to c\bar c q$ topological amplitudes in $\Delta S=-1$ and $\Delta S=0$ transitions in case 1 and 2.
The first uncertainties come from the fitted parameters, while the second uncertainties come from other parameters.
Note that contributions from $b\to u\bar u q$ transitions are sub-leading in $\Delta S=-1$ transitions, but are more prominent in $\Delta S=0$ transitions.
The enhancement can be understood by using the double CKM ratio shown in Eqs. (\ref{eq: 20}) and (\ref{eq: 20 1}).
We expect that the $b\to u\bar u q$ contributions do not significantly alter the situation in $\Delta S=-1$ transitions, but may have large impacts on the $\Delta S=0$ sector.

The branching ratios of the low-lying
$\overline B_q\to \B_c({\bf \bar 3_f})\overline \B_c ({\bf 3_f})$ decays using the parameters in Table~\ref{tab: c1} for case 1 and case 2
are show in Table \ref{tab: Br BtoBcBcbar 3bar3}.
These results are compared to the results from ref. \cite{Chua:2026awd}, where only $b\to c\bar c q$ contributions are considered.
We see that, for $\Delta S=-1$ transitions, these results are similar.
This can be easily understood by using Eqs.~(\ref{eq: b c}) and (\ref{eq: b a}).
The contributions from $b\to u\bar u q$ transitions in $\Delta S=-1$ transitions are CKM-suppressed.
Hence, including $b\to u \bar u s$ contributions does not significantly alter the results of ref. \cite{Chua:2026awd} in $\Delta S=-1$ transitions.

On the other hand, both case 1 and case 2 improve the results of ref. \cite{Chua:2026awd} in $\Delta S=-1$ transitions, 
by lowering the $B^-\to \Xi_c^0 \bar \Lambda_c^{-}$ rate and raising the $\overline B{}^0\to \Xi_c^+ \bar \Lambda_c^{-}$ rate, resulting better agreement with data.
This is precisely why both cases prefer not-so-small $a_{1,v}$ and $d_1$, as they need to overcome the severe CKM suppression, which can be inferred by comparing Eq.~(\ref{eq: b a}) to Eq. (\ref{eq: b c}) again. 

The situation is more interesting in $\Delta S=0$ transitions. 
As shown in Table~\ref{tab: ratios}, the topological amplitudes from $b\to u\bar u q$ contributions are generally not so small in both cases.
We expect to see some enhancement in rates for modes containing these $b\to u\bar u q$ contributions.
Nevertheless the $\overline B{}^0\to \Lambda_c^+ \bar \Lambda_c^{-}$ decay rate is not enhanced.
This can be understood as the destructive interference effect of $D'_1$ and $F'_1$ mentioned previously.
Of course, the destructive interference is required by the data in the first place.
However, to be able to have such a destructive interference, one needs at least two topological amplitudes in these $b\to u\bar u q$ contributions.
It is thus interesting that the $\overline B{}^0\to \Lambda_c^+ \bar \Lambda_c^{-}$ mode is precisely this case.
Hence, the smallness of the $\overline B{}^0\to \Lambda_c^+ \bar \Lambda_c^{-}$ rate not necessary points to small $b\to u\bar u q$ contributions.

To identify modes that can have large enhancements, one should look at modes that only have single topological amplitudes from the $b\to u\bar u d$ contributions.
From Table~\ref{tab: c1}, we see that $B^-\to \Xi_c^0 \bar \Xi_c^{-}$,
$\overline B{}_s^0\to \Lambda_c^+ \bar \Xi_c^{-}$
and                                      
$\overline B{}^0\to \Xi_c^+ \bar \Xi_c^{-}$ decays are precisely these modes, as they only have $A'_{1,c}$, $D'_{1,s}$ and $F'_{1,c}$ from the $b\to u\bar u d$ contributions, respectively.
No destructive interference among $b\to u\bar u d$ contributions is possible in these modes.

Indeed, we see from Table \ref{tab: Br BtoBcBcbar 3bar3} that, 
the rates of
$B^-\to \Xi_c^0 \bar \Xi_c^{-}$,
$\overline B{}_s^0\to \Lambda_c^+ \bar \Xi_c^{-}$
and                                      
$\overline B{}^0\to \Xi_c^+ \bar \Xi_c^{-}$ decays in both cases,
can be much larger than those reported in ref. \cite{Chua:2026awd}.
It will be interesting to search for these three modes to clarify the role of these $b\to u\bar u d$ contributions.

\subsection{Comments on other modes}

From the above analysis, we see that including $b\to u\bar u s$ contributions basically does not change the results in $\Delta S=-1$ transitions.
Hence, previous results in all other modes in $\Delta S=-1$ transitions studied in ref. \cite{Chua:2026awd} can still be useful.
However, the predictions on $\Delta S=0$ transitions may be subject to significant changes.
It will be interesting to keep an eye on $\Delta S=0$ modes in Tables \ref{tab: BtoBcBcbar II} to \ref{tab: BtoBcBcbar IV} which have $b\to u\bar u d$ contributions.
Indeed, it is interesting to note in $\overline B_q\to \B_c({\bf 6_f})\overline \B_c ({\bf 3_f})$ decays
and $\overline B_q\to \B_c({\bf \bar 3_f})\overline \B_c ({\bf \bar 6_f})$ decays that,
$\overline B{}_s^0\to \Sigma_c^+ \bar \Lambda_c^{-}$, $\overline B{}^0\to \Xi_c^{\prime +} \bar \Xi_c^{-}$,
$\overline B{}_s^0\to \Lambda_c^+ \bar \Sigma_c^{-}$ and $\overline B{}^0\to \Xi_c^+ \bar \Xi_c^{\prime -}$ decays, 
only have $W$-exchange diagrams from $b\to u\bar u q$ contributions, see Tables \ref{tab: BtoBcBcbar II} and \ref{tab: BtoBcBcbar III}.
They are sensitive probes of $b\to u\bar u q$ contributions, at least for the $W$-exchange diagram contributions.
As there are no measurements on $\Delta S=0$ modes reported so far, 
except for the low-lying $\overline B_q\to \B_c({\bf \bar 3_f})\overline \B_c ({\bf 3_f})$ modes previously discussed, 
we do not proceed further on $\overline B_q\to \B_c({\bf 6_f})\overline \B_c ({\bf 3_f})$ decays
and $\overline B_q\to \B_c({\bf \bar 3_f})\overline \B_c ({\bf \bar 6_f})$ decays and $\overline B_q\to \B_c({\bf \bar 3_f})\overline \B_c ({\bf 3_f})$ decays with exited charmed baryons.

\begin{table}[t!]
\caption{\label{tab: Br BtoBcBcbar IV}
Some $\overline B_q\to \B_c({\bf 6_f})\overline \B_c ({\bf \bar 6_f})$ decay amplitudes and branching ratios (in units of $10^{-4}$) in $\Delta S=-1$ and $\Delta S=0$ transitions.
Contributions from $b\to u\,\bar u s$ in $\Delta S=-1$ and $\Delta S=0$ transitions are neglected. Modes with $E^{(\prime)}_4$ contributions are not predictable. 
}
\footnotesize{
\begin{ruledtabular}
\begin{tabular}{lccc}
Mode
          & $A \left(\overline B_q\to \B_c({\bf 6_f})\overline \B_c ({\bf \bar 6_f})\right)$
          & $Br \left(\overline B_q\to \B_c({\bf 6_f})\overline \B_c ({\bf \bar 6_f})\right)$
          & Expt.
          \\
\hline
           $B^-\to \Xi_c^{\prime +} \bar \Sigma_c^{--}$
          & $\sqrt2 (C_{4,v}+D_{4,v}+A_{4,v})$ 
          & $14.87^{+4.09}_{-3.80}$
          & $16.8\pm3.1\pm 1.2^{+14.9}_{-5.4}$ \cite{Belle-II:2026ynf}
\\
          $\overline B{}^0\to \Xi_c^{\prime 0} \bar \Sigma_c^{0}$
          & $\sqrt2 C_{4,v}$
          & $13.77^{+3.79}_{-3.52}$
          & $12.8\pm 3.2\pm 1.0^{+3.0}_{-2.1}$ \cite{Belle-II:2026ynf}
\\
$B^-\to \Xi_c^{\prime 0} \bar \Sigma_c^{-}$
          & $C_{4,v}+A_{4,v}$ 
          & $7.45^{+2.05}_{-1.90}$
          & $-$
\\     
$\overline B{}^0\to \Xi_c^{\prime +} \bar \Sigma_c^{-}$
          & $C_{4,v}+D_{4,v}$
          & $6.91^{+1.90}_{-1.77}$
          & $-$
\\
$B^-\to \Omega_c^{0} \bar \Xi_c^{\prime -}$
          & $\sqrt2 (C_{4,vc}+A_{4,vc})$
          & $2.33_{-2.13}^{+5.91}$ 
          & $-$
\\
$\overline B{}^0\to \Omega_c^{0} \bar \Xi_c^{\prime 0}$
          & $\sqrt2 C_{4,vc}$
          & $2.10_{-1.93}^{+5.34}$
          & $-$
\\
\hline
         $B^-\to \Sigma_c^+ \bar \Sigma_c^{- -}$
          & $\sqrt2 (C'_4+D'_4+A'_4)$
          & $0.97^{+2.46}_{-0.60}$
          & $-$
\\
$B^-\to \Sigma_c^0 \bar \Sigma_c^{ -}$
          & $\sqrt2 (C'_4+A'_4)$
          & $0.97^{+2.46}_{-0.60}$
          & $-$
          \\
$B^-\to \Xi_c^{\prime 0} \bar \Xi_c^{\prime -}$
          & $C'_{4,c}+A'_{4,c}$
          & $0.28^{+0.71}_{-0.17}$
          & $-$
          \\
 $\overline B{}_s^0\to \Sigma_c^+ \bar \Xi_c^{\prime -}$
          & $C'_{4,s}+D'_{4,s}$
          & $0.43^{+1.09}_{-0.27}$
          & $-$
          \\                                         
$\overline B{}_s^0\to \Sigma_c^0 \bar \Xi_c^{\prime 0}$
          & $\sqrt2 C'_{4,s}$
          & $0.85^{+5.09}_{-0.74}$
          & $-$
          \\
$\overline B{}_s^0\to \Xi_c^{\prime 0} \bar \Omega_c^{0}$
          & $\sqrt2 C'_{4,cs}$
          & $0.46^{+2.72}_{-0.39}$
          & $-$
          \\                         
\end{tabular}
\end{ruledtabular}
}
\end{table}

We shall comment on $\overline B_q\to \B_c({\bf 6_f})\overline \B_c ({\bf \bar 6_f})$ modes as there are some new experimental results.
Belle II reported the $B^-\to \Xi_c^{\prime +}\bar\Sigma_c(2455)^{--}$ and $\bar B^0\to \Xi_c^{\prime 0}\bar\Sigma_c(2455)^0$ rates
recently~\cite{Belle-II:2026ynf}. 
It will be interesting to find out the implications for other $\overline B_q\to \B_c({\bf 6_f})\overline \B_c ({\bf \bar 6_f})$ modes.
In Table~\ref{tab: Br BtoBcBcbar IV}, by fitting to data the estimation on some $\overline B_q\to \B_c({\bf 6_f})\overline \B_c ({\bf \bar 6_f})$ decay rates in $\Delta S=-1$ and $\Delta S=0$ transitions are given.
Contributions from $b\to u \bar u s$ are neglected.
As we discussed previously, this should be a good approximation.
Modes with $E_4$ contributions are not predictable and are not shown in the table.
We obtain $c_{4,v}=0.290^{+0.038}_{-0.040}$ from the fit.
Its value is similar to $c_{1,v}$ in the $\overline B\to {\cal B}_c(\bf {\bar 3_f}) \overline {\cal B}_c(\bf { 3_f})$ decays.
The uncertainties are from the uncertainty in $c_{4,v}$ and by taking $\delta^{c_4}_v=0\pm 0.4$ and $\delta^{c_4}_s=0\pm 0.4$.

For  $\Delta S=0$ transition modes,
the last two modes, 
namely $\overline B{}_s^0\to \Sigma_c^0 \bar \Xi_c^{\prime 0}$
and $\overline B{}_s^0\to \Xi_c^{\prime 0} \bar \Omega_c^{0}$ decays, 
do not have any $b\to u\bar u d$ contributions, as their amplitudes only consist of $C'_4$.
Their rates can also be estimated. 
For other $\Delta S=0$ modes, we only consider $b\to c\,\bar c d$ contributions. These modes can be used to check the impact of $b\to u\bar u d$ contributions by comparing their values to data in the future. 
If sizable $b\to u\bar u d$ contributions exist in $\overline B_q\to \B_c({\bf 6_f})\overline \B_c ({\bf \bar 6_f})$ decays, their rates should differ noticeably from those in the table, as we see in the $\overline B\to {\cal B}_c(\bf {\bar 3_f}) \overline {\cal B}_c(\bf { 3_f})$ case.  
It will be interesting to measure them.

\section{Conclusion}

The study of decay rates of two-body charmed anti-charmed baryonic $\overline B\to {\cal B}_c \overline {\cal B}_c$ decays using the topological amplitude approach is revisited.
We include the $b\to u\bar u q$ contributions, in addition to the $b\to c\bar c q$ contributions, to $\overline B\to {\cal B}_c \overline {\cal B}_c$ decays.
Our findings are as follows.
\begin{itemize}

\item
Including $b\to u\bar u q$ contributions, the topological decomposition of $\overline B\to {\cal B}_c(\bf {\bar 3_f}) \overline {\cal B}_c(\bf { 3_f})$,
${\cal B}_c(\bf 6_f) \overline {\cal B}_c(\bf { 3_f})$,
${\cal B}_c(\bf {\bar 3_f}) \overline {\cal B}_c(\bf {\bar 6_f})$
and
${\cal B}_c(\bf 6_f) \overline {\cal B}_c(\bf {\bar 6_f})$ decay amplitudes are updated accordingly, see Tables \ref{tab: BtoBcBcbar I}, \ref{tab: BtoBcBcbar II}, \ref{tab: BtoBcBcbar III} and \ref{tab: BtoBcBcbar IV}. 

\item
Although the $b\to u\bar u q$ contributions are CKM suppressed in $\Delta S=-1$ transitions, their effects are significantly amplified in $\Delta S=0$ transitions, 
as the relative size of the CKM factors in $b\to u\bar u q$ and $b\to c\bar c q$ contributions in $\Delta S=0$ transitions is enlarged by $\lambda^{-2}$, roughly a factor of 20, from the one in $\Delta S=-1$ transitions, see Eqs. (\ref{eq: 20}) and (\ref{eq: 20 1}).

\item
We consider two cases in $\overline B\to {\cal B}_c(\bf {\bar 3_f}) \overline {\cal B}_c(\bf { 3_f})$ decays with $b\to c\bar c q$ parameters as input.
The agreement in $B^-\to \Xi_c^0 \bar \Lambda_c^{-}$ and $\overline B{}^0\to \Xi_c^+ \bar \Lambda_c^{-}$ decay rates to data are improved and
the smallness of the $\overline B{}^0\to \Lambda_c^+ \bar \Lambda_c^{-}$ rate not necessary points to small $b\to u\bar u q$ contributions.

\item 
As shown in Table~\ref{tab: Br BtoBcBcbar 3bar3}, these $b\to u\bar u q$ contributions allow much larger rates for $B^-\to \Xi_c^0 \bar \Xi_c^{-}$,
$\overline B{}_s^0\to \Lambda_c^+ \bar \Xi_c^{-}$
and                                      
$\overline B{}^0\to \Xi_c^+ \bar \Xi_c^{-}$ decays than those only have $b\to c\bar c d$ contributions. 
Measuring these decays will be interesting and useful in clarifying the role of $b\to u\bar u q$ contributions in two-body charmed anti-charmed baryonic $\overline B\to {\cal B}_c \overline {\cal B}_c$ decays.

\item 
The inclusion of $b\to u\bar u q$ contributions does not have a large impact on the rates in $\Delta S=-1$ transitions.
Therefore, results in these sectors from previous studies are still useful.

\item In $\overline B_q\to \B_c({\bf 6_f})\overline \B_c ({\bf 3_f})$ decays
and $\overline B_q\to \B_c({\bf \bar 3_f})\overline \B_c ({\bf \bar 6_f})$ decays,
$\overline B{}_s^0\to \Sigma_c^+ \bar \Lambda_c^{-}$, $\overline B{}^0\to \Xi_c^{\prime +} \bar \Xi_c^{-}$,
$\overline B{}_s^0\to \Lambda_c^+ \bar \Sigma_c^{-}$ and $\overline B{}^0\to \Xi_c^+ \bar \Xi_c^{\prime -}$ decays, 
only have $W$-exchange diagrams from $b\to u\bar u q$ contributions, see Tables \ref{tab: BtoBcBcbar II} and \ref{tab: BtoBcBcbar III},.
They are sensitive probes of $b\to u\bar u q$ contributions.

\item
Assuming vanishing $b\to u\bar u q$ contributions, rates of several $\overline B\to{\cal B}_c(\bf 6_f) \overline {\cal B}_c(\bf {\bar 6_f})$ modes
are predicted using data reported recently in ref. \cite{Belle-II:2026ynf}. 
Several modes in $\Delta S=0$ transitions can be used to check the impact of $b\to u\bar u d$ contributions when data are available.

\end{itemize}

Note that we do not advocate large or even huge $b\to u\bar u q$ contributions, but point out that these contributions were overlooked and should be considered systematically.
It is interesting that current data allow or even slightly prefers such contributions. 
As their effects are more prominent in the $\Delta S=0$ transitions, there are good places to search for them.

\begin{acknowledgments}
The author thanks Zan Ren for discussions.
This work is supported in part by the National Science and Technology Council of R.O.C.
under Grant Nos. NSTC-114-2112-M-033-002 and NSTC 115-2112-M-033-001.
\end{acknowledgments}


\end{document}